\documentclass[onecolumn,12pt,journal,draftclsnofoot]{IEEEtran}
\usepackage[latin9]{inputenc}
\usepackage{color}
\usepackage{float}
\usepackage{amsmath}
\usepackage{amssymb}
\usepackage{graphicx}

\makeatletter

\providecommand{\tabularnewline}{\\}
\floatstyle{ruled}
\newfloat{algorithm}{tbp}{loa}
\providecommand{\algorithmname}{Algorithm}
\floatname{algorithm}{\protect\algorithmname}

\usepackage{amsfonts}
\usepackage{mathrsfs}
\usepackage{mathrsfs}\usepackage[noblocks]{authblk}
\usepackage[compress]{cite}
\usepackage{bm}
\usepackage{algorithm}
\usepackage{algorithmic}
\usepackage{epsfig}\usepackage{graphics}\usepackage{subfigure}\usepackage{epsfig}\usepackage{epstopdf}
\usepackage{graphics}\usepackage{subfigure}\usepackage{upgreek}\usepackage{caption}\usepackage{theorem}
\usepackage{breqn}
\usepackage{multicol}
\usepackage{eufrak}
\usepackage{eucal}
\usepackage{array}
\usepackage[compress]{cite}
\usepackage{algorithm}
\usepackage{algorithmic}
\allowdisplaybreaks[4]

\newtheorem{theorem}{Theorem}\newtheorem{lemma}{Lemma}\theoremheaderfont{\normalfont\bfseries}

\makeatother

\begin{document}
\title{Fairness-Oriented Multiple RIS-Aided mmWave Transmission: Stochastic
Optimization Methods}
\author{Gui~Zhou, Cunhua~Pan, Hong~Ren, Kezhi~Wang, and Marco Di Renzo,~\IEEEmembership{Fellow, IEEE}
 \thanks{(Corresponding author: Cunhua Pan) G. Zhou and C. Pan are with the
School of Electronic Engineering and Computer Science at Queen Mary
University of London, London E1 4NS, U.K. (e-mail: g.zhou, c.pan@qmul.ac.uk).
H. Ren is with the National Mobile Communications Research Laboratory,
Southeast University, Nanjing 210096, China. (hren@seu.edu.cn). K.
Wang is with Department of Computer and Information Sciences, Northumbria
University, UK. (e-mail: kezhi.wang@northumbria.ac.uk). M. Di Renzo
is with Université Paris-Saclay, CNRS and CentraleSupélec, Laboratoire
des Signaux et Systèmes, Gif-sur-Yvette, France. (e-mail: marco.di-renzo@universite-paris-saclay.fr).}}
\maketitle
\begin{abstract}
In millimeter wave (mmWave) systems, it is challenging to ensure the
reliable connectivity of communication due to the high sensitivity
to the presence of blockages. In order to improve the robustness of
mmWave systems under the presence of random blockages, multiple reconfigurable
intelligent surfaces (RISs) are deployed to enhance the spatial diversity
gain, and robust beamforming is then designed based on a stochastic
optimization for minimizing the maximum outage probability among multiple
users to ensure fairness. Under the stochastic optimization framework,
we adopt the stochastic majorization--minimization (SMM) method and
the stochastic successive convex approximation (SSCA) method to construct
deterministic surrogate problems at each iteration for new channel
realizations, and to obtain closed-form solutions of the precoding
matrix at the base station (BS) and the passive beamforming vectors
at the RISs. Both stochastic optimization methods are proved to converge
to the set of stationary points of the original stochastic problems.
Finally, simulation results show that the proposed robust beamforming
in RIS-aided systems can effectively compensate for the performance
loss caused by the presence of random blockages, especially when the
blockage probability is high, compared with benchmark solutions. 
\end{abstract}


\section{Introduction}

Millimeter wave (mmWave) communication is expected to be a promising
technology to meet the growing demand for data rate in current and
future wireless networks due to the abundant spectrum available at
high frequencies. High frequency communication inevitably has severe
attenuation, but this can be compensated by an array of reasonable
size containing a large number of antennas due to the small wavelength
\cite{mmWave2013}. In addition, the high-directional beams of large
arrays are capable of mitigating the inter-user interference. However,
mmWave communication systems suffer from high penetration loss \cite{VR-2018,VR-2019,blockage2019}.
Hence, mmWave systems are much more susceptible to
the presence of spatial blockages than sub-6 GHz systems, and the
reliability of the communication links for the entire network cannot
always be guaranteed \cite{VR-2018,VR-2019,blockage2019}.

In particular, spatial blockages can be divided into static blockages
(e.g., buildings and other immobile fixtures), dynamic blockages (e.g.,
humans, vehicles, or moving obstructions) and self-blockages (e.g.,
hand blocking of the user itself and blockage from other body parts).
Some statistical models are established to characterize
the properties of the random dynamic blockages and self-blockages
\cite{Bai-blockage,VR-2019,blockage2019}. The authors
of \cite{mmWave-channel} have developed a distance-dependent blockage
probability model, in which the probability of link blocking increases
exponentially with the link length. Furthermore, \cite{blockage-robust1}
and \cite{blockageprobability-2020} proposed a mechanism to predict
the blockage probability via machine learning. When the blockage
probability is obtained, some robust beamforming design strategies
have been proposed in the recent literature \cite{blockage-robust2,blockage-robust3}
to address channel uncertainties caused by the presence of random
blockages. Specifically, the authors of \cite{blockage-robust2} proposed
a worst-case robust beamforming design for application to coordinated
multipoint (CoMP) systems in which all possible combinations of blockage
patterns were considered. Due to the high computational complexity
of the ergodic method in \cite{blockage-robust2}, an outage-minimum
strategy based on a stochastic optimization method was proposed for
CoMP systems in \cite{blockage-robust3} to improve the robustness
of mmWave systems. The idea of adopting multiple base stations (BSs)
in CoMP systems is effective to compensate for the performance loss
caused by the presence of random blockages by exploiting spatial diversity
gains. However, this will incur excessive hardware cost and power
consumption. Another promising scheme proposed in \cite{Gui-tvt}
is to deploy cost-efficient reconfigurable intelligent surfaces (RISs)
in mmWave systems to create an alternative communication link via
the RISs.

Due to the promising advantages in terms of providing energy- and
spectral-efficient communications, RISs have attracted extensive research
attention from both academia and industry \cite{Marco-5,Marco-6,qingqing2019,xianghao-2020}.
An RIS is a thin surface consisting of nearly-passive and reconfigurable
reflecting elements, which reflects the impinging radio waves without
adopting radio frequency (RF) chains. The passive elements on the
RIS can be tuned to alter their electromagnetic response such that
the signals reflected from an RIS can be constructively superimposed
to enhance the signal power at the intended receiver or destructively
combined to avoid the information leakage to undesired receivers.
These characteristics make RISs appealing to be applied in various
communication systems as shown in \cite{Pan-mag}. For instance, RISs
can be applied in single-cell multiple-input and multiple-output (MIMO)
systems \cite{Pan2019intelleget,Baitong2019,Marco-10,Gui-letter,kangda-2020},
multicell MIMO communications \cite{Pan2019multicell}, simultaneous
wireless information and power transfer (SWIPT) systems \cite{Pan2019intelleget,qingqing2020},
secure communications \cite{sheng2020}, mmWave systems \cite{JSAC-mmWave-RIS,Boya,TVT-mmwave-RIS,Marco-11}
and THz systems \cite{Thz-cunhua,ning-ICC}.

Although the performance advantages of deploying an RIS in mmWave
systems have been demonstrated in recent contributions, there still
exist major open problems to solve. The authors of \cite{Boya} only
considered the BS-RIS-user channels and assumed that the direct BS-user
communication links were completely blocked by obstacles. However,
this assumption only applies to the case of static blockages with
outage probability of 1, but not to the case of dynamic blockage with
outage probability varying between 0 and 1 \cite{VR-2018,VR-2019,blockage2019}.
The numerical results in \cite{TVT-mmwave-RIS} showed that the gain
from additional reflection channels could compensate for the performance
loss caused by the presence of random blockages, but the impact of
blockages was not considered in the beamforming design. Most recently,
we have considered the robust beamforming design for RIS-aided mmWave
communication systems in \cite{Gui-tvt} by taking the random blockages
into consideration. However, the objective function therein is to
minimize the sum outage probability, which cannot ensure the fairness
for all the users.

\subsection{Novelty and contributions}

Against the above background, this paper proposes a robust transmission
strategy in an RIS-aided mmWave communication system to deal with
the channel uncertainties caused by the random blockages while ensuring
the fairness among the users. Some common approaches to handle the
imperfect and partial CSI problem are the outage constrained robust
optimization and the worst-case robust optimization techniques \cite{GuiTSProbust}.
However, both of them still need the estimation of the instantaneous
CSI, and the worst-case robust method is conservative and hence suboptimal
due to the low probability of occurrence for the worst case. An alternative
approach is to design the robust beamforming by optimizing the statistical
performance under a stochastic optimization framework in which only
the long-term CSI is required. In what follows, we
propose a maximum outage probability minimization problem to be solved
by using the stochastic optimization framework.

Specifically, the main contributions of this work are summarized as
follows: 
\begin{itemize}
\item To the best of our knowledge, this is the first work exploring the
robust beamforming design for an RIS-aided downlink multiuser mmWave
system with the knowledge of long-term CSI and blockage
probability. Specifically, our optimization objective in this work
is to minimize the maximum outage probability of all the users. Different
from the sum outage probability minimization problem in \cite{blockage-robust3,Gui-tvt},
the min-max outage probability objective can ensure the quality of
service (QoS) performance for the worst-case user in the downlink
multiuser system. Due to the non-differentiable objective function,
the stochastic gradient descent (SGD) method adopted in \cite{blockage-robust3,Gui-tvt}
cannot be directly applied. To resolve this problem, two more general
and powerful stochastic optimization frameworks are adopted to jointly
optimize the active precoding at the BS and the passive beamforming
at the RISs. 
\item The single user case is firstly considered to obtain insights for
the robust beamforming design by minimizing the outage probability
in the presence of long-term CSI and random blockages.
Due to the non-deterministic expression of the probability function
in the objective function, we replace it with an expectation function
over the long-term CSI and blockage probability.
Then, the inner step function in the expectation is approximated by
a smooth function which is twice differentiable for the precoding
at the BS and the passive beamforming at the RISs, respectively. The
resulting expectation optimization problem is solved by adopting the
stochastic majorization--minimization (SMM) method in which an upper
bound surrogate function of the original differentiable function is
constructed for a new channel realization at each iteration. The constructed
surrogate problem has closed-form solutions and is computationally
efficient. We prove that the proposed SMM method is guaranteed to
converge to the set of stationary points of the original expectation
minimization problem. 
\item The robust beamforming for a more general multiuser case is designed
by solving a maximum outage probability minimization problem. To tackle
the non-differentiability of the max objective function, we firstly
replace it with its log-sum-exp upper bound. The stochastic successive
convex approximation (SSCA) method is then adopted, which has more
flexibility in the choice of the surrogate function than the SMM method,
and generates closed-form solutions at each iteration. Furthermore,
we prove that the final solution generated by the iterative algorithm
is guaranteed to converge to the stationary point of the original
expectation minimization problem. 
\item We demonstrate through numerical results that the proposed robust
beamforming algorithm outperforms its non-robust counterpart and the
robust beamforming in conventional RIS-free systems both in terms
of maximum outage probability and minimum effective rate if the blockage
probability is high. Moreover, deploying multiple small-size RISs
is shown to be more effective than deploying a single large-size RIS
in terms of improving the performance of the worst-case user. 
\end{itemize}
\,\,\,\,\,\,\,The remainder of this paper is organized as follows.
Section II introduces the system model. The outage probability minimization
problem is formulated for a single-user case in Section III. Section
IV further investigates the min-max outage probability problem for
the multiuser system. Finally, Sections VI and VII report the numerical
results and conclusions, respectively.

\noindent \textbf{Notations:} The following mathematical notations
and symbols are used throughout this paper. Vectors and matrices are
denoted by boldface lowercase letters and boldface uppercase letters,
respectively. The symbols $\mathbf{X}^{*}$, $\mathbf{X}^{\mathrm{T}}$,
$\mathbf{X}^{\mathrm{H}}$, and $||\mathbf{X}||_{F}$ denote the conjugate,
transpose, Hermitian (conjugate transpose), Frobenius norm of matrix
$\mathbf{X}$, respectively. The symbol $||\mathbf{x}||_{2}$ denotes
2-norm of vector $\mathbf{x}$. The symbols $\mathrm{Tr}\{\cdot\}$,
$\mathrm{Re}\{\cdot\}$, $|\cdot|$, $\lambda(\cdot)$, and $\angle\left(\cdot\right)$
denote the trace, real part, modulus, eigenvalue, and angle of a complex
number, respectively. $\mathrm{diag}(\mathbf{x})$ is a diagonal matrix
with the entries of $\mathbf{x}$ on its main diagonal. $[\mathbf{x}]_{m}$
means the $m$-th element of the vector $\mathbf{x}$. The Kronecker
product between two matrices $\mathbf{X}$ and $\mathbf{Y}$ is denoted
by $\mathbf{X}\otimes\mathbf{Y}$. $\mathbf{X}\succeq\mathbf{Y}$means
that $\mathbf{X}-\mathbf{Y}$ is positive semidefinite. Additionally,
the symbol $\mathbb{C}$ denotes complex field, $\mathbb{R}$ represents
real field, and $\mathrm{j}\triangleq\sqrt{-1}$ is the imaginary
unit. The inner product $\left\langle \bullet,\bullet\right\rangle :\mathbb{C}^{M\times N}\times\mathbb{C}^{M\times N}\rightarrow\mathbb{R}$
is defined as $\left\langle \mathbf{X},\mathbf{Y}\right\rangle =\mathbb{R}\{\mathrm{Tr}\{\mathbf{X}^{\mathrm{H}}\mathbf{Y}\}\}.$

\section{System Model}

\subsection{Signal Model}

\begin{figure}
\centering \includegraphics[width=3in,height=2.2in]{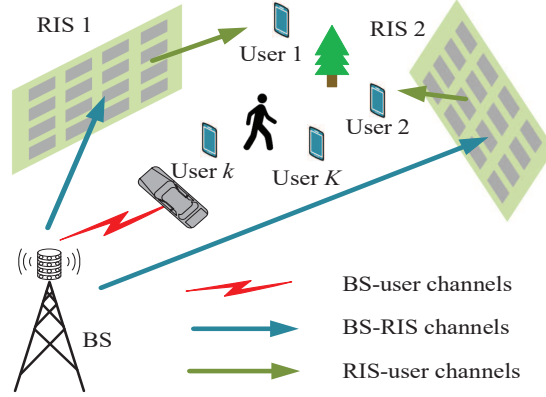}
\caption{Multiple RIS-aided mmWave communication system.}
\label{simulated-model} 
\end{figure}

As shown in Fig. \ref{simulated-model}, we consider an RIS-aided
downlink mmWave communication system. In order to ensure high QoS
for the users in the presence of random blockages, $U$ RISs, each
of which has $M$ passive antennas, are deployed to assist the communication
from the BS equipped with $N$ active antennas to $K$ single-antenna
users (denoted by $\mathcal{K}\triangleq\{1,\ldots,K\}$). The RISs
are assumed to be connected to controllers that exchange control information
with the BS through dedicated channels \cite{Pan2019intelleget,Pan2019multicell}.
The baseband transmitted signal at the BS is $\mathbf{x}=\mathbf{F}{\bf s}$,
where $\mathbf{s}\in\mathbb{C}^{K\times1}\sim\mathcal{CN}(\mathbf{0},\mathbf{I})$
is the Gaussian data symbol vector and $\mathbf{F}=[{\bf \mathbf{f}}_{1},\ldots,{\bf \mathbf{f}}_{K}]\in\mathbb{C}^{N\times K}$
denotes the full-digital beamforming matrix. The
baseband transmit power is limited to the total transmit power $P_{max}$.
Hence, $\mathbf{F}$ belongs to the set $\mathcal{S}_{f}=\{\mathbf{F}\mid||\mathbf{F}||_{F}^{2}\leq P_{max}\}$.

Let $\mathbf{h}_{\mathrm{b},k}\in\mathbb{C}^{N\times1}$, $\mathbf{H}_{u}\in\mathbb{C}^{M\times N}$
and $\mathbf{h}_{u,k}\in\mathbb{C}^{M\times1}$ denote the channels
of the BS-user $k$, BS-RIS $u$, and RIS $u$-user $k$ links, respectively.
Then, the received signal intended to the $k$-th user is expressed
as 
\begin{align}
y_{k} & =\left(\mathbf{h}_{\mathrm{b},k}^{\mathrm{H}}+\sum_{u=1}^{U}\mathbf{h}_{u,k}^{\mathrm{H}}\mathbf{E}_{u}\mathbf{H}_{u}\right)\mathbf{x}+n_{k},\label{eq:1}
\end{align}
where $n_{k}\sim\mathcal{CN}(0,\sigma_{k}^{2})$ is the additive white
Gaussian noise (AWGN), and $\mathbf{E}_{u}=\zeta\mathrm{diag}([e_{(u-1)M+1},\ldots,e_{uM}])$
is the reflection coefficient matrix (also known as the passive beamforming
matrix) of the $u$-th RIS. Element $e_{(u-1)M+m}$ is the unit modulus
coefficient of the $m$-th phase shift at the $u$-th RIS, and $\zeta\in[0,1]$
indicates the reflection efficiency. Here, it is assumed that $\zeta=1$
for investigation of the performance upper bound of RIS.

Furthermore, after defining the compact matrices $\mathbf{h}_{k}=[\mathbf{h}_{1,k}^{\mathrm{H}},\ldots,\mathbf{h}_{U,k}^{\mathrm{H}}]^{\mathrm{H}}$
and $\mathbf{H}=[\mathbf{H}_{1}^{\mathrm{H}},\ldots,\mathbf{H}_{U}^{\mathrm{H}}]^{\mathrm{H}}$,
we obtain the equivalent channel $\mathbf{G}_{k}=\left[\begin{array}{c}
\mathrm{diag}(\mathbf{h}_{k}^{\mathrm{H}})\mathbf{H}\\
\mathbf{h}_{\mathrm{b},k}^{\mathrm{H}}
\end{array}\right]\in\mathbb{C}^{(UM+1)\times N}$ between the BS and the $k$-th user. The corresponding equivalent
reflection coefficient vector is given by $\mathbf{e}=[e_{1},\ldots,e_{UM+1}]^{\mathrm{T}}\in\mathbb{C}^{(UM+1)\times1}$
which belongs to the set $\mathcal{S}_{e}=\{\mathbf{e}\mid|e_{m}|^{2}=1,1\leq m\leq UM,e_{UM+1}=1\}$.
Then, (\ref{eq:1}) can be rewritten into a compact form as 
\begin{equation}
y_{k}=\mathbf{e}^{\mathrm{H}}\mathbf{G}_{k}\mathbf{F}{\bf s}+n_{k},
\end{equation}
and the corresponding achievable signal-to-interference-plus-noise
ratio (SINR) is 
\begin{equation}
\Gamma_{k}\left(\mathbf{F},\mathbf{e}\right)=\frac{|\mathbf{e}^{\mathrm{H}}\mathbf{G}_{k}{\bf f}_{k}|^{2}}{\sum_{i\neq k}^{K}|\mathbf{e}^{\mathrm{H}}\mathbf{G}_{k}{\bf f}_{i}|^{2}+\sigma_{k}^{2}}.\label{eq:Rate-k-1}
\end{equation}

\subsection{Channel Model}

It is important to mention that the perfect instantaneous CSI of a
mmWave system is difficult to obtain due to the passive implementation
of the RIS reflecting elements. More specifically, estimating the
full instantaneous CSI for all links would inevitably require a large
training overhead. In this work, since the channel
modeling of the RIS-related links is the same as the traditional wireless
channel modeling \cite{shanpu}, we adopt a Saleh-Valenzuela
(SV) channel model \cite{SV-MODEL} to characterize the mmWave channels
where only the large-scale fading characteristics is required during
the transmission design. In particular, it is assumed that the uniform
planar array (UPA) is deployed on the BS and each RIS. The steering
vector of the UPA is $\mathbf{a}\left(\varphi,\phi\right)$ in which
$\varphi(\phi)$ denotes the azimuth (elevation) AoD and the angle-of-arrival
(AoA) of the transceivers. We assume that there are $L_{\mathrm{b},k}$,
$L_{u,k}$ and $L_{\mathrm{b},u}$ sparse scatterers for the BS-user
$k$, the RIS $u$-user $k$ and the BS-RIS $u$ links, respectively,
and each of the scatterers comprises $I$ subpaths. Thus, the mmWave
channels can be expressed as 
\begin{align}
\mathbf{h}_{\mathrm{b},k} & =g_{0}^{\mathrm{b},k}\mathbf{a}\left(\varphi_{\mathrm{b},k,0}^{\mathrm{AoD}},\phi_{\mathrm{b},k,0}^{\mathrm{AoD}}\right) +\sqrt{\frac{1}{IL_{\mathrm{b},k}}}\sum_{l=1}^{L_{\mathrm{b},k}}\sum_{i=1}^{I}g_{l,i}^{\mathrm{b},k}\mathbf{a}\left(\varphi_{\mathrm{b},k,l,i}^{\mathrm{AoD}},\phi_{\mathrm{b},k,l,i}^{\mathrm{AoD}}\right),\forall k,\label{eq:3}\\
\mathbf{h}_{u,k} & =g_{0}^{u,k}\mathbf{a}\left(\varphi_{u,k,0}^{\mathrm{AoD}},\phi_{u,k,0}^{\mathrm{AoD}}\right) +\sqrt{\frac{1}{IL_{u,k}}}\sum_{l=1}^{L_{u,k}}\sum_{i=1}^{I}g_{l,i}^{u,k}\mathbf{a}\left(\varphi_{u,k,l,i}^{\mathrm{AoD}},\phi_{u,k,l,i}^{\mathrm{AoD}}\right),\forall k,\forall u,\label{eq:4}\\
\mathbf{H}_{u} & =g_{0}^{\mathrm{b},u}\mathbf{a}\left(\varphi_{u,0}^{\mathrm{AoA}},\phi_{u,0}^{\mathrm{AoA}}\right)\mathbf{a}\left(\varphi_{\mathrm{b},0}^{\mathrm{AoD}},\phi_{\mathrm{b},0}^{\mathrm{AoD}}\right)^{\mathrm{H}}\nonumber \\
&\ \ +  \sqrt{\frac{1}{IL_{\mathrm{b},u}}}\sum_{l=1}^{L_{\mathrm{b},u}}\sum_{i=1}^{I}g_{l,i}^{\mathrm{b},u}\mathbf{a}\left(\varphi_{u,l,i}^{\mathrm{AoA}},\phi_{u,l,i}^{\mathrm{AoA}}\right)\mathbf{a}\left(\varphi_{\mathrm{b},l,i}^{\mathrm{AoD}},\phi_{\mathrm{b},l,i}^{\mathrm{AoD}}\right)^{\mathrm{H}}, \forall u,\label{eq:5}
\end{align}
where, by denoting an arbitrary element $q\in\{(\mathrm{b},k),(u,k),(\mathrm{b},u)\}_{\forall k,\forall u}$,
then $g_{0}^{q}\mathbf{a}\left(\varphi_{q,0}^{\mathrm{AoD}},\phi_{q,0}^{\mathrm{AoD}}\right)$
is the line-of-sight (LoS) component with fading coefficient following
the distribution $g_{0}^{q}\sim\mathcal{CN}(0,\zeta_{0}^{q}10^{\frac{\mathrm{PL}}{10}})$,
where $\zeta_{0}^{q}=\frac{\kappa}{1+\kappa}$ is the power fraction
that corresponds to the Rician factor $\kappa$, and $\mathrm{PL}$
is the distance-dependent large-scale fading model. The remaining
paths are the line-of-sight (NLoS) components whose fading coefficients
follow the distribution $g_{l,i}^{q}\sim\mathcal{CN}(0,\zeta_{l}^{q}10^{\frac{\mathrm{PL}}{10}})$
with power fraction $\zeta_{l}^{q}=\frac{1}{(L_{q}-1)(1+\kappa)}$.

We assume that the user's locations are quasi-static over
milliseconds or even seconds. Therefore, the large-scale fading characteristic
parameters, such as the distance-dependent path loss, the numbers
of clusters, their power fraction, the cluster central angles and
angular beamspreads, change relatively slowly \cite{book-Goldsmith}
and can be perfectly known by the BS. However, the instantaneous CSI,
given by $\{\mathbf{h}_{\mathrm{b},k},\mathbf{h}_{u,k},\mathbf{H}_{u}\}$\footnote{With the knowledge of long-term large-scale fading
characteristic parameters, each sample of $\{\mathbf{h}_{\mathrm{b},k},\mathbf{h}_{u,k},\mathbf{H}_{u}\}$
can be obtained by generating $\{g_{0}^{q},g_{l,i}^{q}\}$, AoDs and
AoAs according to their Gaussian distribution.}, vary during the transmission due to the rapidly varying small-scale
fading coefficients $\{g_{0}^{q},g_{l,i}^{q}\}$, AoDs and AoAs according
to an ergodic stationary process. These AoDs and AoAs can be generated
according to a Gaussian distribution, whose expectation values are
the cluster central angles and variance values are the angular spreads
\cite{book-Goldsmith}.

Based on the fact that the communication links in the mmWave frequency
band are sensitive to the presence of blockages, most existing contributions
considered a worst-case scenario where the BS-user links are completely
blocked by the obstacles during the whole transmission, while the
RIS-related links are not affected by blockages since the locations
of the RISs can be appropriately chosen to ensure line-of-sight transmission.
However, this assumption may be impractical in many scenarios. Traditionally,
blockage effects are incorporated into the shadowing model, along
with reflections, scattering, and diffraction \cite{Bai-blockage}.
In contrast, we adopt a recently proposed probabilistic
model \cite{blockage-robust3} to characterize the channel uncertainties
caused by the presence of random blockages. In particular, by introducing
the blockage parameters $\omega_{k,l}\in\{0,1\},0\leq l\leq L_{\mathrm{b},k},\forall k\in\mathcal{K}$,
which are random variables following a Bernoulli distribution with
a blockage probability of $p_{k}$, the channels in (\ref{eq:3})
for the BS-user links are modified to 
\begin{align}
\mathbf{h}_{\mathrm{b},k}= & \omega_{k,0}g_{0}^{\mathrm{b,}k}\mathbf{a}\left(\varphi_{\mathrm{b},k,0}^{\mathrm{AoD}},\phi_{\mathrm{b},k,0}^{\mathrm{AoD}}\right) +\sqrt{\frac{1}{I}}\sum_{l=1}^{L_{\mathrm{b},k}}\omega_{k,l}\sum_{i=1}^{I}g_{l,i}^{\mathrm{b},k}\mathbf{a}\left(\varphi_{\mathrm{b},k,l,i}^{\mathrm{AoD}},\phi_{\mathrm{b},k,l,i}^{\mathrm{AoD}}\right),\forall k,\label{eq:6}
\end{align}
Analytical results show that the probability of link
blocking increases exponentially with the link length \cite{Bai-blockage},
thus we adopt a distance-dependent blockage probability model given
by $p_{k}(d_{k})=\max(0,1-e^{-a_{out}d_{k}+b_{out}})$, where physical
distance parameter $d_{k}$ between the BS and user $k$ can be extracted
from the estimated CSI by the BS, and parameters $a_{out}$ and $b_{out}$
are fit from the data \cite{mmWave-channel}.

\subsection{Problem Formulation}

To account for the presence of random blockages and to enhance the
QoS of the users, we design a robust beamforming by minimizing the
maximum outage probability among all users over the random
 CSI. The formulated optimization problem is given by\begin{subequations}\label{Pro:multi-1}
\begin{align}
\mathop{\min}\limits _{\mathbf{F},\mathbf{e}} & \;\;\max_{k\in\mathcal{K}}\mathrm{Pr}\{\Gamma_{k}\left(\mathbf{F},\mathbf{e}\right)\leq\gamma_{k}\}\label{eq:P-obj}\\
\textrm{s.t.} & \thinspace\thinspace\thinspace\mathbf{F}\in\mathcal{S}_{f}\label{eq:c1-1}\\
 & \thinspace\thinspace\thinspace\mathbf{e}\in\mathcal{S}_{e},\label{eq:c1-2}
\end{align}
\end{subequations}where the outage probability $\mathrm{Pr}\{\Gamma_{k}\left(\mathbf{F},\mathbf{e}\right)\leq\gamma_{k}\}$
is the probability that the SINR $\Gamma_{k}\left(\mathbf{F},\mathbf{e}\right)$
of the user $k$ is less than the SINR reliability threshold $\gamma_{k}$
for all possible realizations of the random channel $\mathbf{G}=[\mathbf{G}_{1},\ldots,\mathbf{G}_{K}]$.

Compared with the sum outage probability minimization problem in \cite{Gui-tvt},
the objective function in (\ref{Pro:multi-1}) can ensure the fairness
among the users. However, due to the min-max function, the objective
function is not smooth or differentiable, and the algorithms in \cite{Gui-tvt}
cannot be directly applied.

\section{Single-User System}

In this section, we consider a single-user system model in order to
obtain some design insights. By setting $K=1$ and dropping the user
index, Problem (\ref{Pro:multi-1}) is simplified to \begin{subequations}\label{Pro:single1}
\begin{align}
\mathop{\min}\limits _{\mathbf{f},\mathbf{e}} & \;\;\mathrm{Pr}\{\Gamma\left(\mathbf{f},\mathbf{e}\right)\leq\gamma\}\label{eq:P-obj-1}\\
\textrm{s.t.} & \thinspace\thinspace\thinspace\mathbf{f}\in\mathcal{S}_{f}\label{eq:c2-1}\\
 & \thinspace\thinspace\thinspace\mathbf{e}\in\mathcal{S}_{e}.\label{eq:c2-2}
\end{align}
\end{subequations}

\subsection{Problem Reformulation}

The probability $\mathrm{Pr}\{\Gamma\left(\mathbf{f},\mathbf{e}\right)\leq\gamma\}$
has no closed-form expression and thus the problem in (\ref{Pro:single1})
is prohibitively challenging to be solved directly. An alternative
solution is to replace the probability function with an equivalent
expectation function, i.e., $\mathrm{Pr}\{\Gamma(\mathbf{f},\mathbf{e})\leq\gamma\}=\mathbb{E}_{\mathbf{G}}[\mathbb{I}_{\Gamma\leq\gamma}]$
where $\mathbb{I}_{\Gamma\leq\gamma}$ is the step function of the
event $\Gamma\leq\gamma$. By doing so, various stochastic programming
techniques can be used. However, the step function is discontinuous,
and the existing stochastic programming methods cannot be directly
applied.

To resolve this issue, we approximate the step function with a smooth
approximating function 
\begin{equation}
u\left(x\right)=\frac{1}{1+e^{-\theta x}},\label{eq:smooth}
\end{equation}
where $x=\gamma-\Gamma$ and $\theta$ is the smooth parameter controlling
the approximation error.

By defining $f\left(\mathbf{f},\mathbf{e}|\mathbf{G}\right)=u\left(\gamma\sigma^{2}-|\mathbf{e}^{\mathrm{H}}\mathbf{G}{\bf f}|^{2}\right)$,
a convenient approximation of Problem (\ref{Pro:single1}) is 
\begin{align}
\mathop{\min}\limits _{\mathbf{f}\in\mathcal{S}_{f},\mathbf{e}\in\mathcal{S}_{e}} & \;\;g(\mathbf{f},\mathbf{e}|\mathbf{G})=\mathbb{E}\left[f\left(\mathbf{f},\mathbf{e}|\mathbf{G}\right)\right].\label{Pro:single2}
\end{align}

\subsection{Stochastic Majorization-Minimization Method}

A simple approach for solving the above problem is the sample average
approximation (SAA) method. However, the SAA method is computationally
prohibitive since it requires large-sized memory storage due to the
fact that the solution obtained at each iteration is calculated by
averaging over a large number of channel realizations. To overcome
these difficulties, we adopt the widely used SMM \cite{SMM} (also
known as stochastic successive minimization \cite{SSM}) method in
which a well-chosen upper bound approximation of the function $f(\mathbf{f},\mathbf{e}|\mathbf{G})$
for a new channel sample is constructed at each iteration and the
solution is obtained on the average of the new channel sample and
the others generated in previous iterations.

The standard of constructing the upper bound approximation
function of $f(\mathbf{f},\mathbf{e}|\mathbf{G})$ is to make the
corresponding surrogate problem easy to solve and even obtain the
closed-form solutions. In particular, denote $\mathbf{x}\in\{\mathbf{f},\mathbf{e}\}$
that belongs to $\mathcal{S}_{x}\in\{\mathcal{S}_{f},\mathcal{S}_{e}$\},
the surrogate function $\hat{f}(\mathbf{x},\mathbf{x}^{i-1}|\mathbf{G})$
of $f(\mathbf{x}|\mathbf{G})$ around any feasible point $\mathbf{x}^{i-1}$
needs to satisfy the following assumptions \cite{SSM}.

\textbf{Assumption A} 
\begin{align*}
\mathrm{(A1):} & \thinspace\thinspace\hat{f}\left(\mathbf{x},\mathbf{x}^{i-1}|\mathbf{G}\right)\thinspace\thinspace\textrm{is continuous in }\ensuremath{\mathbf{x}}\textrm{ for }\ensuremath{\forall}\ensuremath{\mathbf{x}^{i-1}\in\mathcal{S}_{x}}.\\
\mathrm{(A2):} & \thinspace\thinspace\hat{f}(\mathbf{x}^{i-1},\mathbf{x}^{i-1}|\mathbf{G})=f(\mathbf{x}^{i-1}|\mathbf{G}),\forall\mathbf{x}^{i-1}\in\mathcal{S}_{x}.\\
\mathrm{(A3):} & \thinspace\thinspace\hat{f}(\mathbf{x},\mathbf{x}^{i-1}|\mathbf{G})\geq f(\mathbf{x}|\mathbf{G}),\forall\mathbf{x},\mathbf{x}^{i-1}\in\mathcal{S}_{x}.\\
\mathrm{(A4):} & \thinspace\thinspace\hat{f}^{'}(\mathbf{x}^{i-1}|\mathbf{G};\mathbf{d})=f^{'}(\mathbf{x}^{i-1}|\mathbf{G};\mathbf{d}),\textrm{for all }\mathbf{x}^{i-1}\in\mathcal{S}_{x} \textrm{ \textrm{and} feasible directions }\forall\mathbf{d}\textrm{ at }\mathbf{x}^{i-1}.
\end{align*}
$f^{'}(\mathbf{x}^{i-1}|\mathbf{G};\mathbf{d})$ defines the directional
derivative of $f(\mathbf{x}^{i-1}|\mathbf{G})$ in the direction $\mathbf{d}$
and is given by 
\begin{align*}
f^{'}(\mathbf{x}^{i-1}|\mathbf{G};\mathbf{d})=\lim_{\lambda\rightarrow0}\frac{f(\mathbf{x}^{i-1}+\lambda\mathbf{d}|\mathbf{G})-f(\mathbf{x}^{i-1}|\mathbf{G})}{\lambda}.
\end{align*}

The Assumptions (A2)-(A3) indicate that the surrogate function $\hat{f}(\mathbf{x},\mathbf{x}^{i-1}|\mathbf{G})$
is a locally upper bound of the original function $f(\mathbf{x}|\mathbf{G})$
around the feasible point $\mathbf{x}^{i-1}$. Assumption (A4) is
a derivative consistency condition. To ensure the convergence of the
SMM algorithm, we further make the following assumptions \cite{SSM}.

\textbf{Assumption B} 
\begin{align*}
\mathrm{(B1):} & \textrm{ The feasible set }\mathcal{S}_{x}\textrm{ and channel realizations are bounded.}\\
\mathrm{(B2):} & \textrm{ The functions }\ensuremath{\hat{f}}(\ensuremath{\mathbf{x}},\ensuremath{\mathbf{x}^{i-1}}|\ensuremath{\mathbf{G}})\textrm{ and }f(\ensuremath{\mathbf{x}}|\ensuremath{\mathbf{G}})\textrm{, their  derivatives, and their second-order}\\
 & \textrm{  derivatives are uniformly bounded.}
\end{align*}

Since the variables $\mathbf{f}$ and $\mathbf{e}$
are highly coupled with each other, we adopt an alternating optimization
(AO) method to update them. Based on the above assumptions, the variables
$\mathbf{f}$ and $\mathbf{e}$ are updated by solving the following
two SMM subproblems: 
\begin{align}
\mathbf{f}^{n}=\mathrm{arg}\mathop{\min}\limits _{\mathbf{f}\in\mathcal{S}_{f}} & \;\;\frac{1}{n}\sum_{i=1}^{n}\hat{f}\left(\mathbf{f},\mathbf{f}^{i-1}|\mathbf{G}^{i}\right),\label{Pro:singlef}
\end{align}
and 
\begin{align}
\mathbf{e}^{n}=\mathrm{arg}\mathop{\min}\limits _{\mathbf{e}\in\mathcal{S}_{e}} & \;\;\frac{1}{n}\sum_{i=1}^{n}\hat{f}\left(\mathbf{e},\mathbf{e}^{i-1}|\mathbf{G}^{i}\right).\label{Pro:singlee}
\end{align}
Here, $\mathbf{G}^{1},\mathbf{G}^{2},...$ are some independent samples
of the random equivalent channel $\mathbf{G}$. $\hat{f}\left(\mathbf{f},\mathbf{f}^{i-1}|\mathbf{G}^{i}\right)$
is a surrogate function corresponding to $\mathbf{f}$ when $\mathbf{e}$
is given, while $\hat{f}\left(\mathbf{e},\mathbf{e}^{i-1}|\mathbf{G}^{i}\right)$
is the corresponding surrogate function of $\mathbf{e}$ with given
$\mathbf{f}$.

\subsubsection{Optimizing $\mathbf{f}$}

First, we construct $\hat{f}\left(\mathbf{f},\mathbf{f}^{i-1}|\mathbf{G}^{i}\right)$
satisfying Assumptions A and B, which is shown in the following lemma.
\begin{lemma}\label{lemma-f} For the twice differentiable function
$f(\mathbf{f}|\mathbf{G}^{i})$, we construct its second-order upper
bound approximation around any fixed $\mathbf{f}^{i-1}$, which is
given by 
\begin{align}
\hat{f}(\mathbf{f},\mathbf{f}^{i-1}|\mathbf{G}^{i})=2\textrm{\ensuremath{\mathrm{Re}}}\left\{ \mathbf{d}_{f}^{i,\mathrm{H}}\mathbf{f}\right\} +\alpha_{f}^{i}||\mathbf{f}||_{2}^{2}+\textrm{const}_{f}^{i},\label{quadratic-f-3}
\end{align}
where \begin{subequations} 
\begin{align}
 & \mathbf{d}_{f}^{i}=\mathbf{m}_{f}^{i}-\alpha_{f}^{i}\mathbf{f}^{i-1},\label{eq:q1}\\
 & \mathbf{m}_{f}^{i}=\frac{-\theta e^{-\theta x^{i}}}{\left(1+e^{-\theta x^{i}}\right)^{2}}\mathbf{\mathbf{G}}^{i,\mathrm{H}}\mathbf{e}^{i-1}\mathbf{e}^{i-1,\mathrm{H}}\mathbf{G}^{i}\mathbf{f}^{i-1},\label{eq:q2}\\
 & x^{i}=\gamma\sigma^{2}-|\mathbf{e}^{i-1,\mathrm{H}}\mathbf{G}^{i}{\bf f}^{i-1}|^{2},\label{eq:Q5}\\
 & \alpha_{f}^{i}=\frac{\theta^{2}}{2}P_{max}|\mathbf{e}^{i-1,\mathrm{H}}\mathbf{G}^{i}\mathbf{\mathbf{G}}^{i,\mathrm{H}}\mathbf{e}^{i-1}|^{2},\label{eq:q3}\\
 & \textrm{const}_{f}^{i}=f(\mathbf{f}^{i-1}|\mathbf{G}^{i})+\alpha_{f}^{i}||\mathbf{f}^{i-1}||_{2}^{2}-2\textrm{\ensuremath{\mathrm{Re}}}\left\{ \mathbf{m}_{f}^{i,\mathrm{H}}\mathbf{f}^{i-1}\right\} .\label{eq:q4}
\end{align}
\end{subequations}\end{lemma}

\textbf{\textit{Proof: }}See Appendix \ref{subsec:The-proof-1}.\hspace{4.5cm}$\blacksquare$

By using (\ref{quadratic-f-3}) and ignoring constants,
the subproblem in (\ref{Pro:singlef}) for updating $\mathbf{f}$
is formulated as 
\begin{align}
\mathop{\min}\limits _{\mathbf{f}\in\mathcal{S}_{f}} & \;\;2\textrm{\ensuremath{\mathrm{Re}}}\left\{ \frac{1}{n}\sum_{i=1}^{n}\mathbf{d}_{f}^{i,\mathrm{H}}\mathbf{f}\right\} +\frac{1}{n}\sum_{i=1}^{n}\alpha_{f}^{i}||\mathbf{f}||_{2}^{2}.\label{Pro:single4}
\end{align}
 Problem (\ref{Pro:single4}) is convex and can be solved by computing
its Lagrange function given by 
\begin{align}
\mathcal{L}\text{(\ensuremath{\mathbf{f}},\ensuremath{\kappa})} & =2\textrm{\ensuremath{\mathrm{Re}}}\left\{ \frac{1}{n}\sum_{i=1}^{n}\mathbf{d}_{f}^{i,\mathrm{H}}\mathbf{f}\right\} +\frac{1}{n}\sum_{i=1}^{n}\alpha_{f}^{i}||\mathbf{f}||_{2}^{2}+\kappa\left(||\mathbf{f}||_{2}^{2}-P_{max}\right),\label{eq:Lf}
\end{align}
 where $\kappa\geq0$ is a Lagrange multiplier associated with the
power constraint. By setting $\partial\mathcal{L}\text{(\ensuremath{\mathbf{f}})}/\partial\mathbf{f}^{*}=\mathbf{0}$,
the globally optimal solution of $\mathbf{f}$ at the $n$-th iteration
is derived as 
\begin{equation}
\mathbf{f}^{n}=\frac{-1}{\kappa+\frac{1}{n}\sum_{i=1}^{n}\alpha_{f}^{i}}\frac{1}{n}\sum_{i=1}^{n}\mathbf{d}_{f}^{i}.\label{eq:optimal-F-tao}
\end{equation}

Also, (\ref{eq:optimal-F-tao}) must satisfy the power constraint,
which yields 
\begin{equation}
\frac{||\frac{1}{n}\sum_{i=1}^{n}\mathbf{d}_{f}^{i}||_{2}^{2}}{(\kappa+\frac{1}{n}\sum_{i=1}^{n}\alpha_{f}^{i})^{2}}\leq P_{max}.\label{eq:constraint-tao}
\end{equation}
Based on the fact that the left hand side of (\ref{eq:constraint-tao})
is a decreasing function of $\kappa$, we obtain the following closed-form
solution 
\begin{equation}
\mathbf{f}^{n}=\begin{cases}
\frac{-1}{\sum_{i=1}^{n}\alpha_{f}^{i}}\sum_{i=1}^{n}\mathbf{d}_{f}^{i}, & \textrm{if }\frac{||\sum_{i=1}^{n}\mathbf{d}_{f}^{i}||_{2}^{2}}{(\sum_{i=1}^{n}\alpha_{f}^{i})^{2}}\leq P_{max},\\
-\sqrt{\frac{P_{max}}{||\sum_{i=1}^{n}\mathbf{d}_{f}^{i}||_{2}^{2}}}\sum_{i=1}^{n}\mathbf{d}_{f}^{i}, & \textrm{otherwise.}
\end{cases}\label{eq:f-solution}
\end{equation}
The first option in (\ref{eq:f-solution}) is based on $\kappa=0$.
The second option is due to the fact that there must exist a $\kappa>0$
that (\ref{eq:constraint-tao}) holds with equality.

\subsubsection{Optimizing $\mathbf{e}$}

As for the update of $\mathbf{e}$ with given $\mathbf{f}$, we first
construct a surrogate function corresponding to $\mathbf{e}$ in the
following lemma. \begin{lemma}\label{lemma-e} For the twice differentiable
function $f\left(\mathbf{e}|\mathbf{G}^{i}\right)$, we construct
its second-order upper bound approximation around any feasible $\mathbf{e}^{i-1}$,
which is given by 
\begin{align}
\hat{f}(\mathbf{e},\mathbf{e}^{i-1}|\mathbf{G}^{i})=2\textrm{\ensuremath{\mathrm{Re}}}\left\{ \mathbf{d}_{e}^{i,\mathrm{H}}\mathbf{e}\right\} +\textrm{const}_{e}^{i},\label{quadratic-e}
\end{align}
where \begin{subequations} 
\begin{align}
 & \mathbf{d}_{e}^{i}=\mathbf{m}_{e}^{i}-\alpha_{e}^{i}\mathbf{e}^{i-1},\label{eq:U-1-2-1}\\
 & \mathbf{m}_{e}^{i}=\frac{-\theta e^{-\theta x^{i}}}{\left(1+e^{-\theta x^{i}}\right)^{2}}\mathbf{G}^{i}\mathbf{f}^{i-1}\mathbf{f}^{i-1,\mathrm{H}}\mathbf{G}^{i,\mathrm{H}}\mathbf{e}^{i-1},\\
 & \alpha_{e}^{i}=\frac{\theta^{2}}{2}(UM+1)|\mathbf{f}^{i-1,\mathrm{H}}\mathbf{G}^{i,\mathrm{H}}\mathbf{G}^{i}\mathbf{f}^{i-1}|^{2},\\
 & \textrm{const}_{e}^{i}=f(\mathbf{e}^{i-1}|\mathbf{G}^{i})+2(UM+1)\alpha_{e}^{i}-2\textrm{\ensuremath{\mathrm{Re}}}\left\{ \mathbf{m}_{e}^{i,\mathrm{H}}\mathbf{e}^{i-1}\right\} .\label{constant-f-1-2-1}
\end{align}
\end{subequations}\end{lemma}

\textbf{\textit{Proof: }}The proof of Lemma \ref{lemma-e} is similar
to that of Lemma \ref{lemma-f} and hence omitted for brevity.\hspace{4cm}$\blacksquare$

By substituting (\ref{quadratic-e}) into the objective function of
subproblem (\ref{Pro:singlee}) and ignoring constants,
we obtain: 
\begin{align}
\mathop{\min}\limits _{\mathbf{e}\in\mathcal{S}_{e}} & \;\;2\textrm{\ensuremath{\mathrm{Re}}}\left\{ \frac{1}{n}\sum_{i=1}^{n}\mathbf{d}_{e}^{i,\mathrm{H}}\mathbf{e}\right\} .\label{eq:p-e}
\end{align}
 The globally optimal solution of the above problem is given by
\begin{align}
\mathbf{e}^{n} & =\exp\left\{ \mathrm{j}\angle\left(\left(\sum_{i=1}^{n}\mathbf{d}_{e}^{i}\right)/\left[\sum_{i=1}^{n}\mathbf{d}_{e}^{i}\right]_{UM+1}\right)\right\} ,\label{eq:e-solution}
\end{align}
where $[\cdot]_{m}$ means the $m$-th element of vector, $\mathrm{j}\triangleq\sqrt{-1}$
is the imaginary unit, $\angle\left(\cdot\right)$ denotes the angle
of a complex number, and $\exp\left\{ \mathrm{j}\angle\left(\cdot\right)\right\} $
is an element-wise operation.

\subsection{Algorithm Development}

Under the SMM framework, closed-form solutions of
$\mathbf{f}$ in (\ref{eq:f-solution}) and of $\mathbf{e}$ in (\ref{eq:e-solution})
at each iteration are obtained. Such simple closed-form solutions
can greatly reduce the computational complexity. Algorithm \ref{Algorithm-1}
summarizes the proposed robust beamforming design based on the SMM-based
outage probability minimum problem for RIS-aided single-user mmWave
systems in which the BS-user links experience random blockages. The
proposed algorithm is referred to as SMM-OutMin. Note
that the convergence speed of the SMM-based algorithm might be affected
by the tightness of the upper bounds in Lemma \ref{lemma-f} and Lemma
\ref{lemma-e}, thus SQUAREM \cite{varadhan2008SQUAREM} is adopted
to accelerate the SMM-based algorithm. Denote by $F\left(\mathbf{f}^{n}\right)$
and $F\left(\mathbf{e}^{n}\right)$ the objective function values
of Problem (\ref{Pro:single4}) and Problem (\ref{eq:p-e}) in the
$n$-th iteration, respectively. $\mathcal{\mathcal{P_{S}}}(\cdot)$
is an operation mapped to nonlinear constraint set. For the power
constraint set $\mathcal{S}_{f}$, $\mathcal{\mathcal{P_{S}}}(\cdot)$
can be $\mathcal{\mathcal{P_{S}}}(\mathbf{x})=\frac{\left(\mathbf{x}\right)}{||\mathbf{x}||_{2}}||\mathbf{f}_{2}||_{2}$.
For the unit-modulus constraint set $\mathcal{S}_{e}$, $\mathcal{\mathcal{P_{S}}}(\cdot)$
can be element-wise operation, i.e., $\mathcal{\mathcal{P_{S}}}(\mathbf{x})=j\angle(\mathbf{x})$.
Steps 10 to 13 and steps 21 to 24 are used to maintain the convergence
property of the objective function values.

\begin{algorithm}
\caption{SMM-OutMin Algorithm}
\label{Algorithm-1} \begin{algorithmic}[1] \REQUIRE Initialize
$\mathbf{f}^{0}\in\mathcal{S}_{f}$ and $\mathbf{e}^{0}\in\mathcal{S}_{e}$.
Set $n=1$.

\REPEAT

\STATE Obtain the sample channel $\mathbf{G}^{n}$. \STATE Set $\mathbf{e}=\mathbf{e}^{n-1}$.
\STATE Obtain $\mathbf{f}_{1}$ according to (\ref{eq:f-solution})
based on $\mathbf{f}^{n-1}$. \STATE Obtain $\mathbf{f}_{2}$
according to (\ref{eq:f-solution}) based on $\mathbf{f}_{1}$. 
\STATE
$\mathbf{j}_{1}=\mathbf{f}_{1}-\mathbf{f}^{n-1}$.
\STATE $\mathbf{j}_{2}=\mathbf{f}_{2}-\mathbf{f}_{1}-\mathbf{j}_{1}$.
\STATE $\omega=-\frac{||\mathbf{j}_{1}||_{2}}{||\mathbf{j}_{2}||_{2}}$.
\STATE $\mathbf{f}^{n}=-\mathcal{\mathcal{P_{F}}}(\mathbf{f}^{n-1}-2\omega\mathbf{j}_{1}+\omega^{2}\mathbf{j}_{2})$.
\WHILE {$|F(\mathbf{f}^{n})-F(\mathbf{f}^{n-1})|\leq|F(\mathbf{f}^{n-1})-F(\mathbf{f}^{n-2})|$}
\STATE  $\thinspace\thinspace\thinspace\thinspace\thinspace\thinspace\thinspace\omega=(\omega-1)/2$.
\STATE $\thinspace\thinspace\thinspace\thinspace\thinspace\thinspace\mathbf{f}^{n}=-\mathcal{\mathcal{P_{S}}}(\mathbf{f}^{n-1}-2\omega\mathbf{j}_{1}+\omega^{2}\mathbf{j}_{2})$.

\ENDWHILE

\STATE Set $\mathbf{f}=\mathbf{f}^{n}$. 
\STATE Obtain $\mathbf{e}_{1}$ according to (\ref{eq:e-solution}) based on $\mathbf{e}^{n-1}$.
\STATE Obtain $\mathbf{e}_{2}$ according to (\ref{eq:e-solution}) based on $\mathbf{e}_{1}$. 
\STATE $\mathbf{j}_{1}=\mathbf{e}_{1}-\mathbf{e}^{n-1}$.
\STATE $\mathbf{j}_{2}=\mathbf{e}_{2}-\mathbf{e}_{1}-\mathbf{j}_{1}$.
\STATE $\omega=-\frac{||\mathbf{j}_{1}||_{2}}{||\mathbf{j}_{2}||_{2}}$.
\STATE $\mathbf{e}^{n}=-\mathcal{\mathcal{P_{S}}}(\mathbf{e}^{n-1}-2\omega\mathbf{j}_{1}+\omega^{2}\mathbf{j}_{2})$.

\WHILE{ $|F\left(\mathbf{e}^{n}\right)-F\left(\mathbf{e}^{n-1}\right)|\leq|F\left(\mathbf{e}^{n-1}\right)-F\left(\mathbf{e}^{n-2}\right)|$}
\STATE  $\thinspace\thinspace\thinspace\thinspace\thinspace\thinspace\thinspace\omega=(\omega-1)/2$.
\STATE $\thinspace\thinspace\thinspace\thinspace\thinspace\thinspace\thinspace\mathbf{e}^{n}=-\mathcal{\mathcal{P_{S}}}(\mathbf{e}^{n-1}-2\omega\mathbf{j}_{1}+\omega^{2}\mathbf{j}_{2})$.

\ENDWHILE

\STATE $n=n+1$.

\UNTIL $||F\left(\mathbf{f}^{n}\right)-F\left(\mathbf{f}^{n-1}\right)||_{2}\rightarrow0$
and $||F\left(\mathbf{e}^{n}\right)-F\left(\mathbf{e}^{n-1}\right)||_{2}\rightarrow0$.

\end{algorithmic} 
\end{algorithm}

\subsubsection{Convergence analysis}

The convergence of Algorithm \ref{Algorithm-1} is given in the following
theorem. \begin{theorem}\label{theorem-1} Suppose Assumptions A
and B are satisfied. Then the sequence of the solutions obtained in
each iteration of Algorithm \ref{Algorithm-1} converge to the set
of stationary points of Problem (\ref{Pro:single2}) almost surely.
\end{theorem}

\textbf{\textit{Proof: }}See Appendix \ref{subsec:The-proof-2}.\hspace{4.5cm}$\blacksquare$

\subsubsection{Complexity analysis}

The computational complexity for updating $\mathbf{f}^{n}$ and $\mathbf{e}^{n}$
at each iteration mainly depends on the computation of (\ref{eq:f-solution})
and (\ref{eq:e-solution}), respectively. In particular, due to the
update rule in $\{\sum_{i=1}^{n}\alpha_{f}^{i},\sum_{i=1}^{n}\mathbf{d}_{f}^{i},\sum_{i=1}^{n}\mathbf{d}_{e}^{i}$\},
only $\{\alpha_{e}^{n},\mathbf{d}_{f}^{n},\mathbf{d}_{e}^{n}\}$ needs
to be calculated at the $n$-th iteration. Therefore,
the approximate complexity of each iteration is given by $\mathcal{O}(4UMN+12N)$.

\section{Multiuser System}

In this section, we consider the general multiuser setup and solve
Problem (\ref{Pro:multi-1}). Problem (\ref{Pro:multi-1}) is more
challenging than problem (\ref{Pro:single1}) due to the non-differentiable
objective function. Furthermore, the complex objective function complicates
the use of the SMM algorithm, thus we extend the SMM method to a general
algorithm to solve Problem (\ref{Pro:multi-1}).

\subsection{Problem Reformulation}

We first approximate the probability function in the original Problem
(\ref{Pro:multi-1}) by using the expectation of the smooth function
in (\ref{eq:smooth}). By defining $f_{k}\left(\mathbf{F},\mathbf{e}|\mathbf{G}\right)=u\left(\mathbf{e}^{\mathrm{H}}\mathbf{G}_{k}{\bf F}\boldsymbol{\Upsilon}_{k}\mathbf{F}^{\mathrm{H}}\mathbf{G}_{k}^{\mathrm{H}}\mathbf{e}+\gamma_{k}\sigma_{k}^{2}\right)$,
in which $\boldsymbol{\Upsilon}_{k}$ is a diagonal matrix whose diagonal
entry is $\gamma_{k}$ except in the $k$-th diagonal element that
is $-1$, the approximate objective function is $\max_{k\in\mathcal{K}}\mathbb{E}\left[f_{k}\left(\mathbf{F},\mathbf{e}|\mathbf{G}\right)\right]$.
However, the obtained function is still intractable since the maximization
operation couples $f_{k},\forall k$ and the different channel states
due to the expectation operation. This issue motivates us to use the
following Jensen inequality 
\begin{align}
\max_{k\in\mathcal{K}}\mathbb{E}\left[f_{k}\left(\mathbf{F},\mathbf{e}|\mathbf{G}\right)\right]\leq\mathbb{E}\left[\max_{k\in\mathcal{K}}f_{k}\left(\mathbf{F},\mathbf{e}|\mathbf{G}\right)\right],
\end{align}
due to the fact that the max function $\max_{k\in\mathcal{K}}\{x_{1},\ldots,x_{K}\}$
is convex \cite{book-convex}.

Furthermore, the non-differentiable max function, $\max_{k\in\mathcal{K}}f_{k}\left(\mathbf{F},\mathbf{e}|\mathbf{G}\right)$,
is approximated by adopting a smooth log--sum--exp upper-bound \cite{xu2001smoothing}
\begin{align}
 & \max_{k\in\mathcal{K}}f_{k}\left(\mathbf{F},\mathbf{e}|\mathbf{G}\right)\approx F\left(\mathbf{F},\mathbf{e}|\mathbf{G}\right)\nonumber \\
 & =\mu\ln\Bigl(\sum_{k\in\mathcal{K}}\mathrm{exp}\left\{ \frac{1}{\mu}f_{k}\left(\mathbf{F},\mathbf{e}|\mathbf{G}\right)\right\} \Bigr),\label{eq:dd}
\end{align}
where $\mu>0$ is a smoothing parameter satisfying 
\begin{align}
\max_{k\in\mathcal{K}}f_{k}\left(\mathbf{F},\mathbf{e}|\mathbf{G}\right) & \leq F\left(\mathbf{F},\mathbf{e}|\mathbf{G}\right) \leq\max_{k\in\mathcal{K}}f_{k}\left(\mathbf{F},\mathbf{e}|\mathbf{G}\right)+\frac{1}{\mu}\log\left(|\mathcal{K}|\right).
\end{align}

When $\mu$ is chosen appropriately, the smooth approximation of Problem
(\ref{Pro:multi-1}) is approximately reformulated as 
\begin{align}
\mathop{\min}\limits _{\mathbf{F}\in\mathcal{S}_{f},\mathbf{e}\in\mathcal{S}_{e}} & \;\;G\left(\mathbf{F},\mathbf{e}|\mathbf{G}\right)=\mathbb{E}\left[F\left(\mathbf{F},\mathbf{e}|\mathbf{G}\right)\right].\label{Pro:multi2}
\end{align}

\subsection{Stochastic Successive Convex Approximation Method}

Similar to Problem (\ref{Pro:single2}), Problem (\ref{Pro:multi2})
can still be solved by adopting the SMM method. However, the function
$F\left(\mathbf{F},\mathbf{e}|\mathbf{G}\right)$ in (\ref{eq:dd})
is much more complex and its second-order derivative, which is necessary
to construct the upper bound surrogate function of $F\left(\mathbf{F},\mathbf{e}|\mathbf{G}\right)$
as shown in Appendix \ref{subsec:The-proof-1}, is not easy to be
calculated. Furthermore, the coefficient of the second-order term
in the final upper-bound surrogate function of $F\left(\mathbf{F},\mathbf{e}|\mathbf{G}\right)$
($\alpha_{f}^{i}$ in (\ref{quadratic-f-3})) can be quite loose,
resulting in a very slow convergence rate of the SMM algorithm.

Therefore, in this section, we adopt a flexible SSCA
method to address the above issues. The surrogate functions employed
by the SSCA method do not need to be an upper bound of the original
function but they need only to preserve the first-order property of
the original function. Accordingly, the surrogate functions need to
satisfy Assumption B and the following assumptions \cite{Razaviyayn-thiese}.

\textbf{Assumption C} 
\begin{align*}
\mathrm{(C1):}\thinspace\thinspace & \hat{F}\left(\mathbf{x},\mathbf{x}^{i-1}|\mathbf{G}\right)\thinspace\thinspace\textrm{is strongly convex in \ensuremath{\mathbf{x}}\textrm{ for }\ensuremath{\forall}\ensuremath{\mathbf{x}^{i-1}\in\mathcal{S}_{x}}.}\\
\mathrm{(C2):}\thinspace\thinspace & \hat{F}(\mathbf{x}^{i-1},\mathbf{x}^{i-1}|\mathbf{G})=F(\mathbf{x}^{i-1}|\mathbf{G}),\forall\mathbf{x}^{i-1}\in\mathcal{S}_{x}.\\
\mathrm{(C3):}\thinspace\thinspace & \nabla_{\mathbf{x}}\hat{F}(\mathbf{x}^{i-1},\mathbf{x}^{i-1}|\mathbf{G})=\nabla_{\mathbf{x}}F(\mathbf{x}^{i-1}|\mathbf{G}),\forall\mathbf{x},\mathbf{x}^{i-1}\in\mathcal{S}_{x}.
\end{align*}

Assumption C cannot ensure that the sequences of the approximate objective
values are monotonically decreasing at each iteration. Nevertheless,
to guarantee convergence, the variables can be updated by choosing
an appropriate step size at each iteration that yields a sufficient
decrease of the objective value. Based on the above assumptions, we
choose the proximal gradient-like approximation to construct the surrogate
function, which is 
\begin{align}
\hat{F}\left(\mathbf{x},\mathbf{x}^{i-1}|\mathbf{G}\right)= & F\left(\mathbf{x}^{i-1}|\mathbf{G}\right)+\nabla_{\mathbf{x}}F(\mathbf{x}^{i-1}|\mathbf{G})^{\mathrm{T}}(\mathbf{x}-\mathbf{x}^{i-1}) +\frac{\tau^{i}}{2}||\mathbf{x}-\mathbf{x}^{i-1}||^{2},\label{eq:vv-1}
\end{align}
where $\tau^{i}$ can be any positive number.

\subsubsection{Optimizing $\mathbf{f}$}

By using (\ref{eq:vv-1}), we construct a surrogate function as 
\begin{align}
&\hat{F}\left(\mathbf{F},\mathbf{F}^{i-1}|\mathbf{G}\right) \nonumber \\ & =F\left(\mathbf{F}^{i-1}|\mathbf{G}\right)+\mathrm{Tr}\left(\nabla_{\mathbf{F}}F\left(\mathbf{F}^{i-1}|\mathbf{G}\right)^{\mathrm{T}}(\mathbf{F}-\mathbf{F}^{i-1})\right)+\mathrm{Tr}\left(\nabla_{\mathbf{F}^{*}}F\left(\mathbf{F}^{i-1}|\mathbf{G}\right)^{\mathrm{T}}(\mathbf{F}^{*}-\mathbf{F}^{i-1,*})\right)\nonumber \\
 & \thinspace\thinspace\thinspace\thinspace\thinspace\thinspace+\frac{\tau^{i}}{2}||\mathbf{F}-\mathbf{F}^{i-1}||_{F}^{2}\nonumber \\
 & =F\left(\mathbf{F}^{i-1}|\mathbf{G}\right)+2\sum_{k\in\mathcal{K}}l_{k}^{i}\mathrm{Re}\left\{ \mathrm{Tr}\left(\boldsymbol{\Upsilon}_{k}\mathbf{F}^{i-1,\mathrm{H}}\mathbf{G}_{k}^{\mathrm{H}}\mathbf{e}^{i-1}\mathbf{e}^{i-1,\mathrm{H}}\mathbf{G}_{k}(\mathbf{F}-\mathbf{F}^{i-1})\right)\right\} +\frac{\tau^{i}}{2}||\mathbf{F}-\mathbf{F}^{i-1}||_{F}^{2}\nonumber \\
 & =2\textrm{\ensuremath{\mathrm{Re}}}\left\{ \mathrm{Tr}\left(\mathbf{P}_{f}^{i,\mathrm{H}}\mathbf{F}\right)\right\} +\frac{\tau^{i}}{2}||\mathbf{F}||_{F}^{2}+\textrm{cons1}^{i},\label{eq:su-F}
\end{align}
 for $\mathbf{F}$ around the fixed $\mathbf{F}^{i-1}$
when $\mathbf{e}$ is given. 

The parameters in (\ref{eq:su-F}) are as follows \begin{subequations}
\begin{align}
 & \mathbf{P}_{f}^{i}=\mathbf{W}_{f}^{i}-\frac{\tau^{i}}{2}\mathbf{F}^{i-1},\label{eq:U-1-1-1}\\
 & \mathbf{W}_{f}^{i}=\sum_{k\in\mathcal{K}}l_{k}^{i}\mathbf{G}_{k}^{i,\mathrm{H}}\mathbf{e}^{i-1}\mathbf{e}^{i-1,\mathrm{H}}\mathbf{G}_{k}^{i}\mathbf{F}^{i-1}\boldsymbol{\Upsilon}_{k},\\
 & l_{k}^{i}=\frac{\mathrm{exp}\left\{ \frac{1}{\mu}f_{k}\left(\mathbf{F}^{i-1},\mathbf{e}^{i-1}|\mathbf{G}^{i}\right)\right\} }{\sum_{k\in\mathcal{K}}\mathrm{exp}\left\{ \frac{1}{\mu}f_{k}\left(\mathbf{F}^{i-1},\mathbf{e}^{i-1}|\mathbf{G}^{i}\right)\right\} }\frac{\theta e^{-\theta x_{k}^{i}}}{\left(1+e^{-\theta x_{k}^{i}}\right)^{2}},\\
 & x_{k}^{i}=\mathbf{e}^{i-1,\mathrm{H}}\mathbf{G}_{k}^{i}{\bf F}^{i-1}\boldsymbol{\Upsilon}_{k}\mathbf{F}^{i-1,\mathrm{H}}\mathbf{G}_{k}^{i,\mathrm{H}}\mathbf{e}^{i-1}+\gamma_{k}\sigma_{k}^{2},\\
 & \textrm{cons1}^{i}=F\left(\mathbf{F}^{i-1}|\mathbf{G}^{i}\right)+\frac{\tau^{i}}{2}||\mathbf{F}^{i-1}||_{F}^{2}-2\textrm{\ensuremath{\mathrm{Re}}}\left\{ \mathrm{Tr}\left(\mathbf{W}_{f}^{i,\mathrm{H}}\mathbf{F}^{i-1}\right)\right\} .\label{eq:con-f}
\end{align}
\end{subequations}

By using (\ref{eq:su-F}), the subproblem of Problem (\ref{Pro:multi2})
corresponding to $\mathbf{F}$ at the $n$-th iteration is formulated
as 
\begin{align}
\mathop{\min}\limits _{\mathbf{F}\in\mathcal{S}_{F}} & \;\;\frac{1}{n}\sum_{i=1}^{n}\hat{F}\left(\mathbf{F},\mathbf{F}^{i-1}|\mathbf{G}^{i}\right).\label{eq:fin-F}
\end{align}
The method to solve Problem (\ref{eq:fin-F}) is the same as that
of solving Problem (\ref{Pro:single4}), thus we directly provide
the global minimizer of Problem (\ref{eq:fin-F}) as 
\begin{equation}
\widehat{\mathbf{F}}^{n}=\begin{cases}
\frac{-2}{\sum_{i=1}^{n}\tau^{i}}\sum_{i=1}^{n}\mathbf{P}_{f}^{i}, & \textrm{if }\frac{4||\sum_{i=1}^{n}\mathbf{P}_{f}^{i}||_{F}^{2}}{(\sum_{i=1}^{n}\tau^{i})^{2}}\leq P_{max},\\
-\sqrt{\frac{P_{max}}{||\sum_{i=1}^{n}\mathbf{P}_{f}^{i}||_{F}^{2}}}\sum_{i=1}^{n}\mathbf{P}_{f}^{i}, & \textrm{otherwise.}
\end{cases}\label{eq:F-solution}
\end{equation}

\subsubsection{Optimizing $\mathbf{e}$}

Furthermore, with fixed $\mathbf{F}$, the subproblem of Problem (\ref{Pro:multi2})
corresponding to $\mathbf{e}$ at the $n$-th iteration is 
\begin{align}
\mathop{\min}\limits _{\mathbf{e}\in\mathcal{S}_{e}} & \;\;\frac{1}{n}\sum_{i=1}^{n}\hat{F}\left(\mathbf{e},\mathbf{e}^{i-1}|\mathbf{G}^{i}\right),\label{eq:fin-e}
\end{align}
where $\hat{F}\left(\mathbf{e},\mathbf{e}^{i-1}|\mathbf{G}^{i}\right)=2\textrm{\ensuremath{\mathrm{Re}}}\left\{ \mathbf{p}_{e}^{i,\mathrm{H}}\mathbf{e}\right\} +\textrm{cons2}^{i}$,
and \begin{subequations} 
\begin{align}
 & \mathbf{p}_{e}^{i}=\mathbf{w}_{e}^{i}-\frac{\tau^{i}}{2}\mathbf{e}^{i-1},\label{eq:U-1-1}\\
 & \mathbf{w}_{e}^{i}=\sum_{k\in\mathcal{K}}l_{k}\mathbf{G}_{k}\mathbf{F}^{i-1}\boldsymbol{\Upsilon}_{k}\mathbf{F}^{i-1,\mathrm{H}}\mathbf{\mathbf{G}^{\mathrm{H}}}\mathbf{e}^{i-1},\\
 & \textrm{cons2}^{i}=F\left(\mathbf{e}^{i-1}|\mathbf{G}^{i}\right)+\tau^{i}(UM+1)-2\textrm{\ensuremath{\mathrm{Re}}}\left\{ \mathbf{w}_{e}^{i,\mathrm{H}}\mathbf{e}^{i-1}\right\} .\label{constant-f-1-1}
\end{align}
\end{subequations}

Finally, the minimizer of Problem (\ref{eq:fin-e}) is 
\begin{align}
\widehat{\mathbf{e}}^{n} & =\exp\left\{ \mathrm{j}\angle\left(\left(\sum_{i=1}^{n}\mathbf{p}_{e}^{i}\right)/\left[\sum_{i=1}^{n}\mathbf{p}_{e}^{i}\right]_{UM+1}\right)\right\} .\label{eq:e-solution-1}
\end{align}

\subsection{Algorithm development}

The simple closed-form solutions of $\mathbf{f}$
in (\ref{eq:F-solution}) and of $\mathbf{e}$ in
(\ref{eq:e-solution-1}) can greatly reduce the computational
complexity. Algorithm \ref{Algorithm-2} summarizes the proposed
SSCA-based robust beamforming design for RIS-aided multiuser mmWave
systems in which the BS-user links experience random blockages. The
proposed algorithm is referred to as SSCA-OutMin.

\begin{algorithm}
\caption{SSCA-OutMin Algorithm}
\label{Algorithm-2} \begin{algorithmic}[1] \REQUIRE Initialize
$\mathbf{F}^{0}\in\mathcal{S}_{f}$ and $\mathbf{e}^{0}\in\mathcal{S}_{e}$.
Set $n=0$.

\REPEAT

\STATE $n=n+1$.

\STATE Obtain the sample channel $\mathbf{G}^{n}$.

\STATE Calculate $\widehat{\mathbf{F}}^{n}$ according to (\ref{eq:F-solution}).

\STATE Update $\mathbf{F}^{n}=\mathbf{F}^{n-1}+\xi_{f}^{n}\left(\widehat{\mathbf{F}}^{n}-\mathbf{F}^{n-1}\right)$.

\STATE Calculate $\widehat{\mathbf{e}}^{n}$ according to (\ref{eq:e-solution-1}).

\STATE Update $\mathbf{e}^{n}=\mathbf{e}^{n-1}+\xi_{e}^{n}\left(\widehat{\mathbf{e}}^{n}-\mathbf{e}^{n-1}\right)$.

\UNTIL $||\mathbf{F}^{n}-\mathbf{F}^{n-1}||_{F}^{2}\rightarrow0$
and $||\mathbf{e}^{n}-\mathbf{e}^{n-1}||_{2}\rightarrow0$.

\end{algorithmic} 
\end{algorithm}

\subsubsection{Step-size selection}

It is worth noting that the approximation in (\ref{eq:su-F}) has
the same form as that in (\ref{quadratic-f-3}). However, $\tau^{i}$
in the SSCA method can be any positive number, and $\hat{F}\left(\mathbf{F},\mathbf{F}^{i-1}|\mathbf{G}\right)$
might no longer be a global upper bound of $F\left(\mathbf{F}|\mathbf{G}\right)$.
In this case, the step sizes $\xi_{f}^{n}$ and $\xi_{e}^{n}$ need
to be carefully chosen to ensure convergence.

As an example, we take $\xi_{f}^{n}$ to illustrate the update rule,
which is a line-search (also called Armijo step-size) rule: Consider
$\xi_{f}^{0}>0$ and $c_{1,f},c_{2,f}\in(0,1)$. Let
$\xi_{f}^{n}$ be the largest element in $\{\xi_{f}^{0}c_{2,f}^{t}\}_{t=0,1,\ldots}$
such that 
\begin{align}
F\biggl(\mathbf{F}^{n-1}+ & \xi_{f}^{n}\left(\widehat{\mathbf{F}}^{n}-\mathbf{F}^{n-1}\right)\biggr)\leq F(\mathbf{F}^{n-1}) +c_{1,f}\xi_{f}^{n}\mathrm{Tr}\left(\nabla_{\mathbf{F}}F(\mathbf{F}^{n-1})^{\mathrm{T}}\left(\widehat{\mathbf{F}}^{n}-\mathbf{F}^{n-1}\right)\right).
\end{align}

\begin{theorem}\label{theorem-3} If $\{\xi_{f}^{n}\}_{n=1,2,\ldots}$
is chosen according to the line-search rule, then 
\begin{align*}
\lim_{n\rightarrow\infty}||\widehat{\mathbf{F}}^{n}-\mathbf{F}^{n-1}|| & =0.
\end{align*}
\end{theorem}

\textbf{\textit{Proof: }}See Theorem 7 in \cite{thesis-Razaviyayn}.\hspace{10.5cm}$\blacksquare$

\subsubsection{Convergence analysis}

The convergence of Algorithm \ref{Algorithm-2} is given in the following
theorem.

\begin{theorem}\label{theorem-1-1} Suppose Assumptions B and C are
satisfied. Then every limit point of the iterations generated by Algorithm
\ref{Algorithm-2} is a stationary point of Problem (\ref{Pro:multi2})
almost surely. \end{theorem}

\textbf{\textit{Proof: }}See Appendix \ref{subsec:The-proof-4}.\hspace{11.5cm}$\blacksquare$

\subsubsection{Complexity analysis}

The computational complexity for updating $\mathbf{f}^{n}$ and $\mathbf{e}^{n}$
at each iteration mainly depend on the computation of (\ref{eq:F-solution})
and (\ref{eq:e-solution-1}), respectively. In particular, only $\{\mathbf{P}_{f}^{n},\mathbf{p}_{e}^{n}\}$
needs to be calculated at the $n$-th iteration.
Therefore, the approximate complexity of each iteration is given by
$\mathcal{O}((K+2)2UMN+UMK+NK+(N+2)K^{2}+2N)$.

\subsubsection{Initial point}

Problem (\ref{Pro:multi2}) has, in general, multiple local minima
points due to the non-convex unit-modulus constraint and $\mathbf{e}\in\mathcal{S}_{e}$.
The accurate selection of the initial points in Algorithm \ref{Algorithm-2}
plays an important role for the convergence speed and the quality
of the obtained local solution. To that end, we first initialize $\mathbf{e}$
to maximize the minimum equivalent total channel gain, resulting in
the following optimization problem 
\begin{align}
\mathbf{e}^{0}=\mathrm{arg}\mathop{\max}\limits _{\mathbf{e}\in\mathcal{S}_{e}} & \;\;\min_{k\in\mathcal{K}}||\mathbf{e}^{\mathrm{H}}\mathbf{G}_{k}^{0}||_{2}^{2}.\label{eq:d-1}
\end{align}

Problem (\ref{eq:d-1}) can be efficiently solved by using the SDR
method as follows \begin{subequations} 
\begin{align}
\mathop{\max}\limits _{\mathbf{E}} & \;\;t\label{eq:d-1-1}\\
\textrm{s.t.} & \;\;\mathrm{Tr}\{\mathbf{G}_{k}^{0}\mathbf{G}_{k}^{0,\textrm{H}}\mathbf{E}\}\geq t,\forall k\in\mathcal{K}\\
 & \;\;\mathbf{E}\succeq0,\mathrm{rank}(\mathbf{E})=1,[\mathbf{E}]_{m,m}=1,\forall m,
\end{align}
\end{subequations}where $\mathbf{E}=\mathbf{e}\mathbf{e}^{\mathrm{H}}$
and $t$ is an auxiliary variable.

Furthermore, $\mathbf{F}$ is initialized by using the maximum-ratio
transmission (MRT) method as 
\begin{align}
\mathbf{F}^{0}=P_{max}\frac{\mathbf{G}^{0}\mathbf{e}^{0}}{||\mathbf{G}^{0}\mathbf{e}^{0}||}.
\end{align}

\section{Numerical results and discussion}

\begin{figure}
\centering \includegraphics[width=2.2in,height=1.2in]{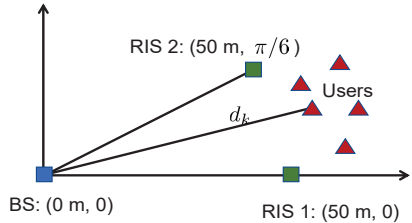}
\caption{The simulated system setup.}
\label{simulation-model} 
\end{figure}

\subsection{Simulation setup}

In this section, we numerically evaluate the performance of the proposed
algorithms. All experiments are performed on a PC with a 1.99 GHz
i7-8550U CPU and 16 GB RAM. We adopt a polar coordinate to describe
the simulated system setup as shown in Fig. \ref{simulation-model},
where the BS is located at (0 m, 0), and the two RISs are deployed
in the locations (50 m, 0) and (50 m, $\pi/6$) which are close to
the users. The users are randomly placed in a range with polar diameter
$d_{k}\in${[}50 m, 80 m{]} and polar angle $\vartheta\in[0,\pi/6]$,
where $d_{k}$ is used to calculate the distance-dependent
blockage probability. The large-scale fading, reported in urban micro
(UMi)-street canyon scenario \cite{3Gpp}, is modeled as $\mathrm{PL}=32.4+20\log_{10}(f_{c})+10\alpha\log_{10}(D)+\xi$
in dB with link distance $D$ (in meters), path loss exponent $\alpha$,
and log-normal shadowing $\xi\sim\mathcal{CN}(0,\sigma_{\mathrm{\xi}}^{2})$
where $\sigma_{\mathrm{\xi}}^{2}$ denotes the log-normal shadowing
variance. The mmWave system operates on carrier frequency $f_{c}=28$
GHz and bandwidth 20 MHz. Since the macro-scattering environment
between the BS and the users are complex, only NLoS clusters are assumed
to exist in the BS-user links, i.e., the Rician factor is $\kappa=0$.
The parameter settings in $\mathrm{PL}$ of NLoS are
$\alpha=3.5$ and $\sigma_{\xi}=8.2$ dB \cite{3Gpp}. In
practice, the RISs can be installed such that the BS-RIS links and
the RIS-user links are blockage-free. Thus, the channels in (\ref{eq:4})
and (\ref{eq:5}) contain only LoS cluster with a Rician factor $\kappa\rightarrow\infty$.
The parameters in $\mathrm{PL}$ of LoS are $\alpha=2$
and $\sigma_{\xi}=4$ dB according to \cite{3Gpp}. Unless stated
otherwise, we assume $L_{\mathrm{b},k}=L_{u,k}=L_{\mathrm{b},u}=5$
and $I=20$. The transmit power limit of the BS is
$P_{max}=30$ dBm and the noise power at each user is $\sigma_{1}^{2}=\ldots=\sigma_{K}^{2}=-94$
dBm. For simplicity, we consider an equal blockage probability, $p_{k,l}=p_{\mathrm{block}},\forall k,l$,
and an equal target SINR, $\gamma=\gamma_{1}=\ldots=\gamma_{K}$,
which yields the target rate $R_{\mathrm{targ}}=\log_{2}(1+\gamma)$.
The smooth parameters are chosen to be $\theta=\frac{1}{\max_{\forall k\in\mathcal{K}}|x_{k}^{0}|}$
and $\mu=\frac{1}{100K}$.

To evaluate the performance of the proposed stochastic optimization
algorithms, we consider the following benchmark schemes. 1) SMRT:
In this scheme, the active precoding is updated by using stochastic
maximum-ratio transmission (SMRT), that is $\mathbf{F}^{n+1}=\frac{\frac{1}{n}\sum_{i=1}^{n}\mathbf{G}^{i,\mathrm{H}}\mathbf{e}^{i}}{||\frac{1}{n}\sum_{i=1}^{n}\mathbf{G}^{i,\mathrm{H}}\mathbf{e}^{i}||_{2}}\sqrt{P_{max}}$.
The passive beamforming is still updated by using the SMM or SSCA
methods. 2) NoRIS: In this case, no RIS is employed and the optimal
active precoding is obtained by using the SMM or SSCA methods. 3)
No-robust: In this scheme, the beamforming is designed by using the
SMM or SSCA methods by taking into account the random small-scale
parameters while assuming $p_{\mathrm{block}}=0$. 4)
Imperfect CSI: In this scheme, the beamforming is designed by using
the SMM or SSCA methods based on the imperfect CSI of cluster central
angles. In specific, we assume that the modulus of the cluster central
angle estimation error is 0.01. 5) SAA: In this scheme, we generate
300 independent channel realizations in advance, the solutions at
each iteration is the average over these 300 channel samples, and
the surrogate function used at each iteration is obtained by adopting
the MM or SCA methods. We take the beamforming design in the single-user
case as an example. By modifying Problems (\ref{Pro:singlef}) and
(\ref{Pro:singlee}), the beamforming designed by using the SAA-MM
method is updated as follows 
\begin{align}
\mathbf{f}^{n}=\mathrm{arg}\mathop{\min}\limits _{\mathbf{f}\in\mathcal{S}_{f}} & \;\;\frac{1}{300}\sum_{i=1}^{300}\hat{f}\left(\mathbf{f},\mathbf{f}^{n-1}|\mathbf{G}^{i}\right),\label{Pro:singlef-1}
\end{align}
and 
\begin{align}
\mathbf{e}^{n}=\mathrm{arg}\mathop{\min}\limits _{\mathbf{e}\in\mathcal{S}_{e}} & \;\;\frac{1}{300}\sum_{i=1}^{300}\hat{f}\left(\mathbf{e},\mathbf{e}^{n-1}|\mathbf{G}^{i}\right).\label{Pro:singlee-1}
\end{align}

In order to demonstrate the robustness of the proposed algorithms,
we consider two performance metrics: the outage probability and the
effective rate. In particular, the outage probability of each user
is calculated by averaging over 1000 independent channel realizations.
The corresponding effective rate of the $k$-th user is defined as
$R_{\mathrm{eff},k}\triangleq\mathbb{E}[\log_{2}(1+\Gamma_{k}(\mathbf{F},\mathbf{e}))]$
if $\Gamma_{k}(\mathbf{F},\mathbf{e})\geq\gamma$ and $R_{\mathrm{eff}}\triangleq0$
otherwise.

\subsection{Convergence}

\begin{figure}
\centering \includegraphics[width=3.2in,height=2.4in]{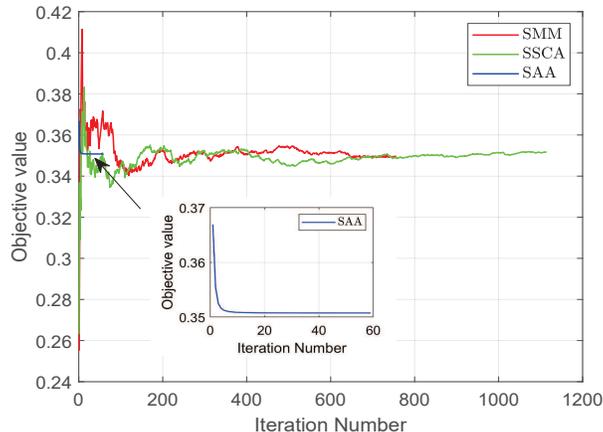}
\caption{Convergence behavior of different algorithms, when
$N=8$, $M=128$, $K=1$, $U=1$, and $R_{\mathrm{targ}}=0.5$ bps/Hz.}
\label{converge} 
\end{figure}

Fig. \ref{converge} investigates the convergence behavior of the
considered stochastic optimization algorithms. For comparison, we
consider a single-user case containing RIS 1 in Fig.
\ref{simulation-model}, and the other parameters are given in Fig.
\ref{converge}. In Fig. \ref{converge}, the coordinate value on
the y-axis is the objective value of Problem (\ref{Pro:single4})
or (\ref{eq:fin-F}), and not the actual outage probability of the
original problem. It is observed from Fig. \ref{converge} that the
SMM and SSCA algorithms are characterized by an oscillatory convergence
which depends on the random channel generations at each iteration.
On the other hand, using 300 channel realizations for each iteration
leads to the monotonic convergence of the SAA algorithm when adopting
a monotonically decreasing surrogate function for each channel realization.
Although the SAA algorithm requires the least number of iterations
to converge, it is much more computationally demanding than the other
two algorithms. This fact can be observed in Table I which compares
the CPU time consumption of each iteration and total
CPU time consumption of iteration convergence for the three considered
algorithms. Theoretically, the computational complexity of each iteration
of the SAA algorithm is 300 times higher than that of the SMM or SAA
algorithms, because at each iteration of the SAA algorithm, each parameter
needs to be calculated 300 times for all channel realizations.

\begin{table}
\centering \label{table} \caption{Comparison of the CPU time}
\begin{tabular}{|c|c|c|}
\hline 
Algorithms  & {The CPU time (sec) per iteration }  & The CPU time (sec)\tabularnewline
\hline 
\hline 
SMM  & {0.0025 }  & 1.8750\tabularnewline
\hline 
SSCA  & {0.0042}  & 4.6719\tabularnewline
\hline 
SAA  & {0.3557 }  & 20.9844\tabularnewline
\hline 
\end{tabular}
\end{table}

\subsection{Single-user system}

\begin{figure}
\centering \subfigure[Outage probability]{ \includegraphics[width=3.2in,height=2.4in]{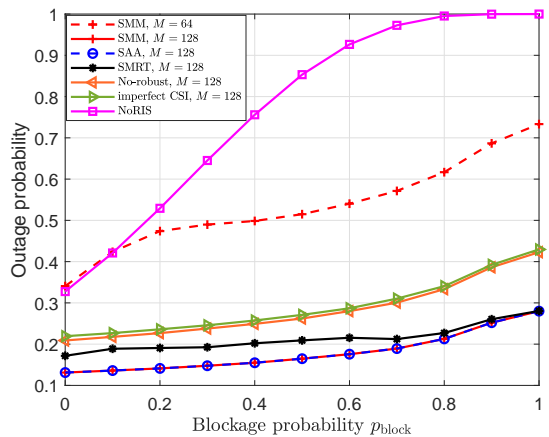}}

\subfigure[Effective rate]{ \includegraphics[width=3.2in,height=2.4in]{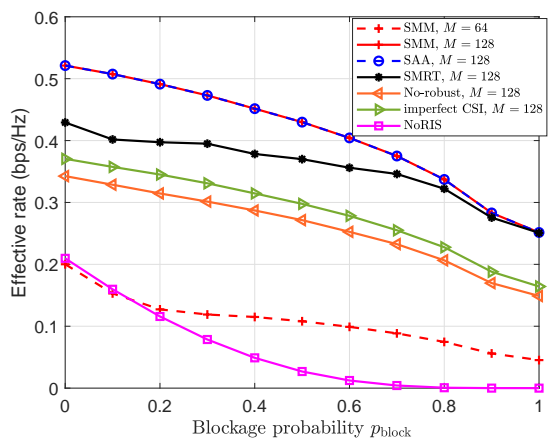}}

\caption{Comparison of outage probability and effective rate
as a function of the blockage probability $p_{\mathrm{block}}$ for
$N=8$, $K=1$, $U=1$, and $R_{\mathrm{targ}}=0.5$ bps/Hz.}
\label{single-blockage} 
\end{figure}

We consider a single-user system containing RIS 1
in Fig. \ref{simulation-model} with the target rate of $R_{\mathrm{targ}}=0.5$
bps/Hz. Fig. \ref{single-blockage} illustrates the performance of
different algorithms as a function of the blockage probability. First,
it can be seen that SMM-based beamforming in the RIS-aided mmWave
system of $M=64$ outperforms the NoRIS scheme when $p_{\mathrm{block}}\geq0.1$.
The main reason is that the direct BS-user channel may be much stronger
than the cascaded BS-RIS-user channel as the latter experiences the
double path loss effect, which is significant if the path-loss is
large. Therefore, when the blockage probability is small ($p_{\mathrm{block}}\leq0.1$),
the BS tends to allocate the transmit power to the stronger direct
path, thus reducing the contribution of the RIS to the system performance.
However, by increasing the number of reflecting elements at the RIS
to $M=128$, our proposed SMM algorithm outperforms the NoRIS system
in the whole blockage probability region (i.e., $0\leq p_{\mathrm{block}}\leq1$).
The reason is that the RIS-aided reflecting channels starts to compensate
for the performance loss caused by the blockages when $p_{\mathrm{block}}=0$.
In addition, it can be seen that the performance of the SMM-based
beamforming is the same as that of the SAA-based beamforming. Last
but not least, compared with the No-robust RIS-aided scenario and
the scheme with imperfect CSI, the robust RIS beamforming can significantly
compensate for the performance loss caused by the presence of random
blockages.

\begin{figure}
\centering \subfigure[Number of reflection elements]{ \includegraphics[width=3.2in,height=2.4in]{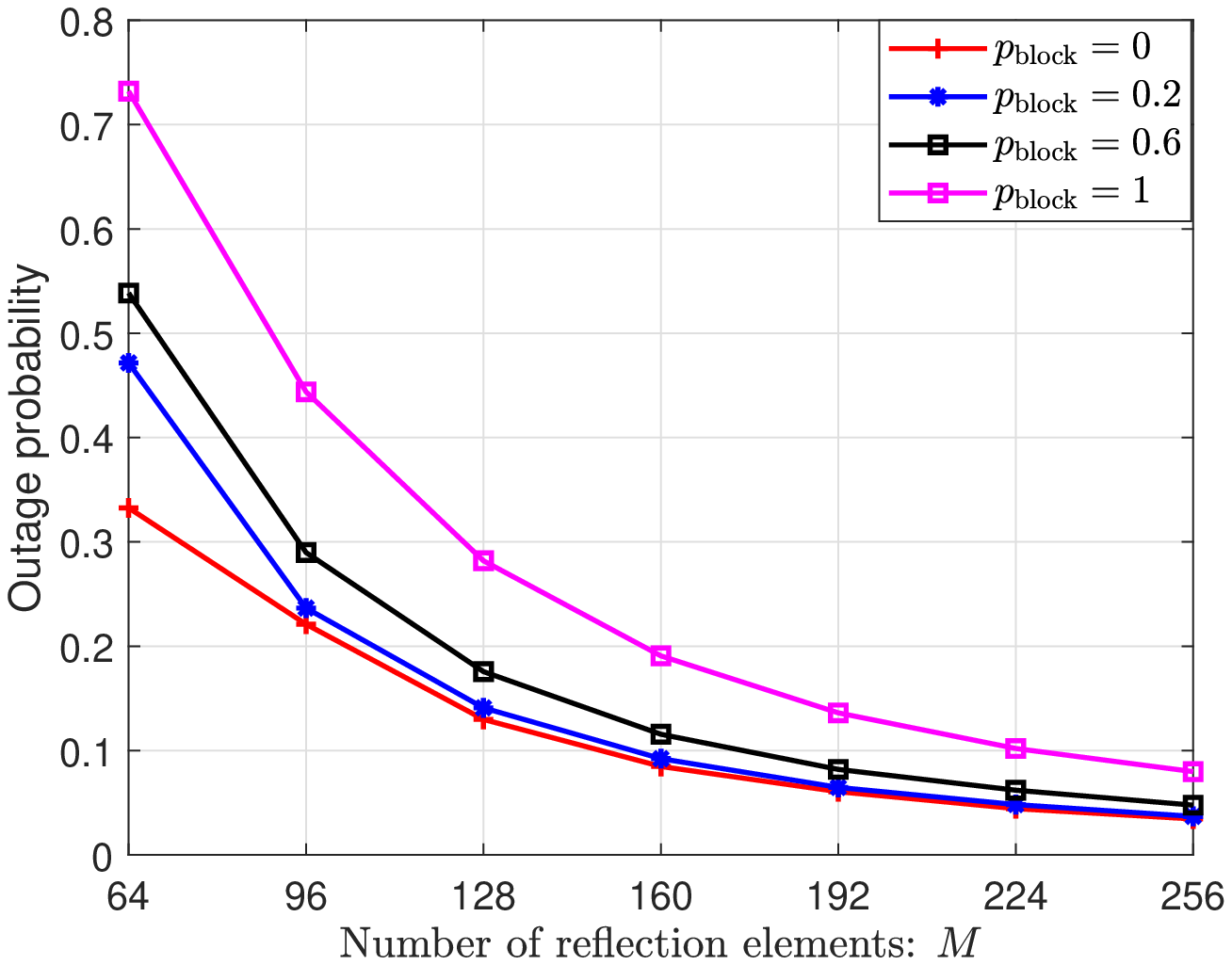}}

\subfigure[Number of transmit antennas]{ \includegraphics[width=3.2in,height=2.4in]{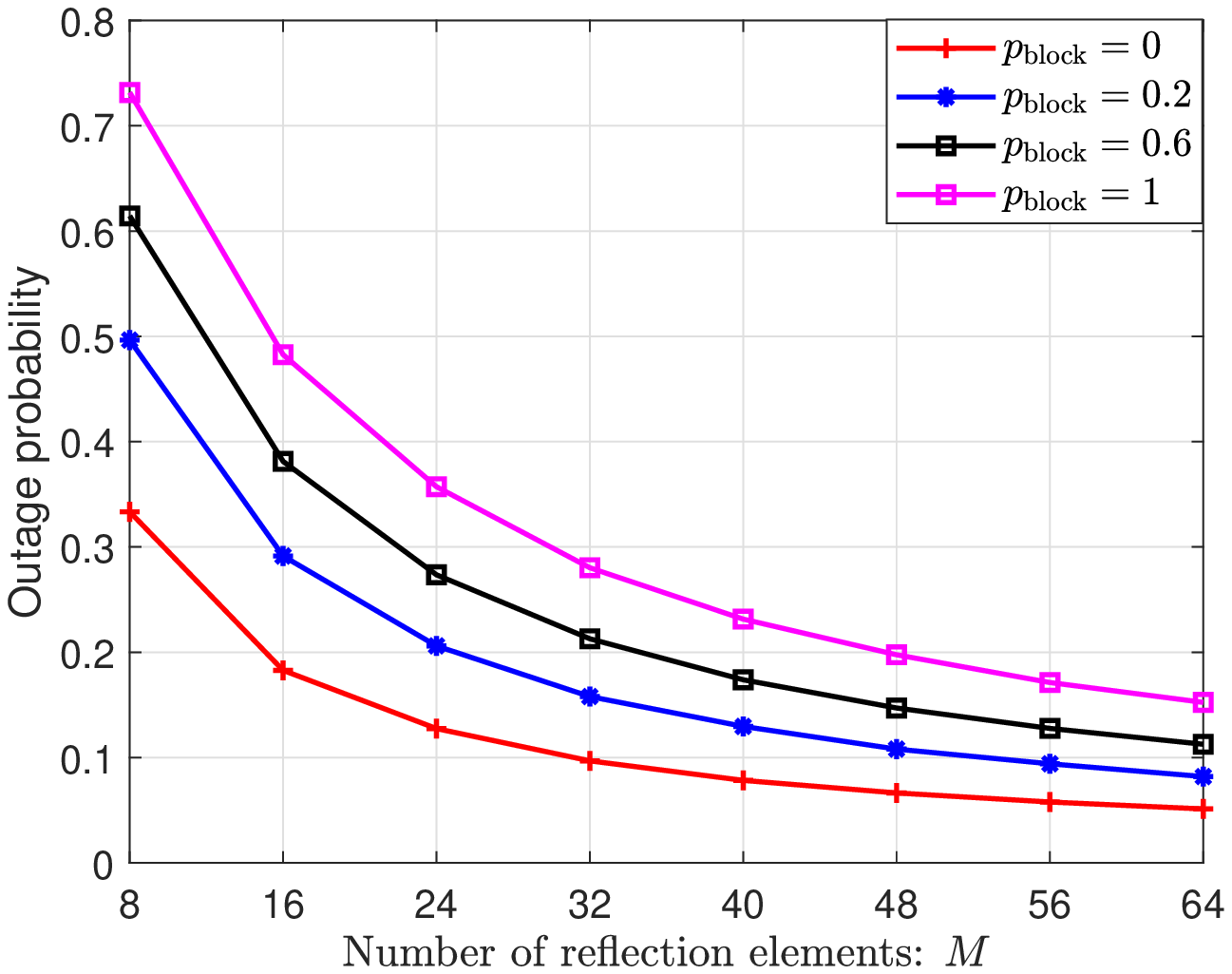}
}

\caption{Comparison of outage probability as a function of
$M$ with fixed $N=8$ and of $N$ with fixed $M=128$, when $K=1$,
$U=1$, and $R_{\mathrm{targ}}=0.5$ bps/Hz.}
\label{single-NM} 
\end{figure}

Fig. \ref{single-NM} shows the impact of the size of the RIS and
the size of the antenna array at the BS on the outage probability
by using the SMM algorithm. It can be observed from
Fig. \ref{single-NM}(a) that when BS is equipped with $N=8$ antennas,
the RIS plays a significant role in guaranteeing the desired user's
QoS and improving the system robustness with increased number of reflection
elements ($M:64\rightarrow256$). In other words, a large-size RIS
with $M\geq224$ can still reduce the outage probability to about
0.1, even if the direct channel is completely blocked. Finally in
Fig. \ref{single-NM}(b) we observe that increasing the number of
transmit antennas ($N:8\rightarrow64$) has the same ability to reduce
the outage probability as increasing the number of reflection elements
($M:64\rightarrow256$), except when the direct channel is completely
blocked.

\subsection{Multiuser system}

\begin{figure}
\centering \subfigure[Outage probability]{ \includegraphics[width=3.2in,height=2.4in]{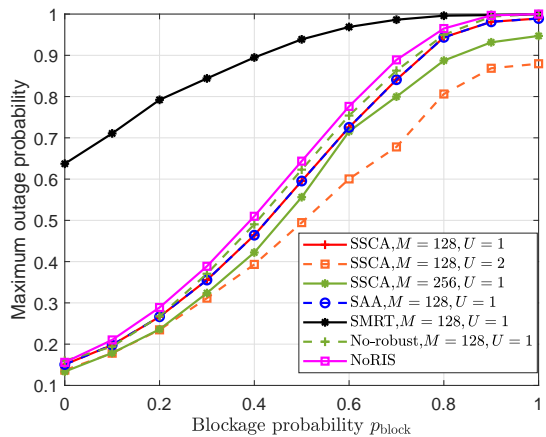}
}

\subfigure[Effective rate]{ \includegraphics[width=3.2in,height=2.4in]{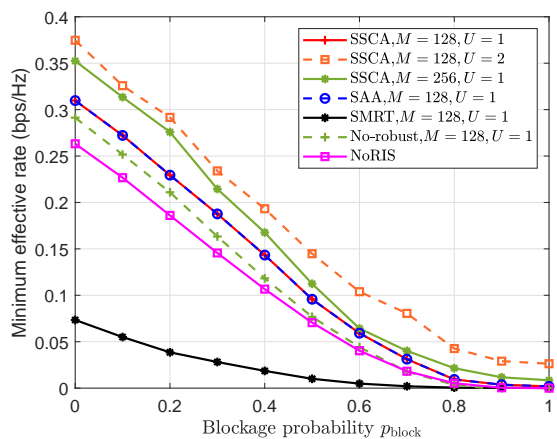}}

\caption{Comparison of maximum outage probability and minimum
effective rate as a function of the blockage probability $p_{\mathrm{block}}$
for $N=16$, $K=3$, and $R_{\mathrm{targ}}=0.1$ bps/Hz.}
\label{multi-blockage} 
\end{figure}

In this section, we consider a multiuser system with
$K=3$ users, in which the target rate is $R_{\mathrm{targ}}=0.1$
bps/Hz. As we can see from Fig. \ref{multi-blockage}(a), compared
with the No-robust RIS-aided scenario, the robust beamforming in the
RIS-aided scheme in a multiuser system can efficiently improve the
maximum outage probability. In addition, different from the single-user
system of $M=128$, the RIS-aided setup outperforms
the NoRIS scheme only when $p_{\mathrm{block}}\geq0.2$. This fact
is consistent with the observation of $M=64$ in Fig. \ref{single-blockage}
(a), that is, the performance gain of an RIS starts to manifest when
$p_{\mathrm{block}}=0.2$. Further increasing the number of RISs and
the size of each RIS can reduce the maximum outage probability in
the whole blockage probability region (i.e., $0\leq p_{\mathrm{block}}\leq1$).
Although the improvement in the performance in terms of the maximum
outage probability of the large-size RIS beamforming in Fig. \ref{multi-blockage}(a)
is slightly weak, the enhancement of the corresponding minimum effective
rate shown in Fig. \ref{multi-blockage}(b) is apparent. Specifically,
the minimum effective rate of the robust designs is better than that
of the non-robust schemes. Moreover, the contribution of the RIS in
improving the minimum effective rate is also apparent for the almost
blockage probability region (i.e., $p_{\mathrm{block}}\leq0.8$).
Finally, we observe that multiple RISs with a relatively small size
can improve the worst-case user's performance, and outperform the
performance offered by a single RIS with large size, especially in
the multi-user scenario. This is because the optimal deployment of
multiple RISs can provide more spatial diversity for users who are
far away from the RIS, thus improving the system performance.

\begin{figure}
\centering \includegraphics[width=3.2in,height=2.4in]{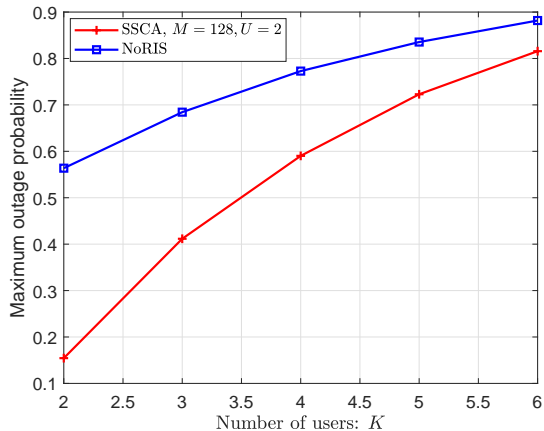}
\caption{Comparison of maximum outage probability as a function
of the number of users $K$ for $N=32$, $p_{\mathrm{block}}=0.6$,
and $R_{\mathrm{targ}}=0.1$ bps/Hz.}
\label{multi-K} 
\end{figure}

Finally, Fig. \ref{multi-K} investigates the performance as a function
of the number of users. For a fair comparison, we consider the setup
$N=32,M=128$ and $U=2$. Compared with the NoRIS scheme, the contribution
of multiple RISs in reducing the maximum outage probability shown
in Fig. \ref{multi-K} decreases as the increase of $K$. When $K\geq5$,
the RIS scheme cannot guarantee the QoS performance of the worst-case
user due to the high outage probability.

\section{Conclusions}

In this work, we have improved the reliability of a mmWave system
in the presence of random blockages by employing multiple RISs and
designing the corresponding robust beamforming. In order to reduce
the system outage, we have formulated and solved a maximum outage
probability minimization problem which belongs to the family of stochastic
optimization problems . More precisely, we have taken into account
the statistical CSI and the blockage probability, and have solved
the formulated optimization problem by adopting a stochastic optimization
framework. Closed-form solutions have been derived at each iteration
by adopting the SMM and SSCA methods. The two proposed stochastic
methods are guaranteed to converge to the set of stationary points
of the original stochastic problems. Selected numerical results have
demonstrated the performance gain in terms of outage probability and
effective rate of the RIS-aided mmWave systems in the presence of
random blockages.

\appendices{}

\section{The proof of Lemma \ref{lemma-f}\label{subsec:The-proof-1}}

In this subsection, $\mathbf{G}^{i}$ is dropped for simplicity, i.e.,
$f\left(\mathbf{f}|\mathbf{G}^{i}\right)$ is replaced by $f\left(\mathbf{f}\right)$.
Since $f\left(\mathbf{f}\right)$ is twice differentiable, we propose
a second-order approximation to upper bound $f\left(\mathbf{f}\right)$
at any fixed point $\mathbf{f}^{i-1}$: 
\begin{align}
f\left(\mathbf{f}\right)\leq & \hat{f}\left(\mathbf{f},\mathbf{f}^{i-1}\right)\nonumber \\
= & f\left(\mathbf{f}^{i-1}\right)+2\mathrm{Re}\left\{ \mathbf{m}_{f}^{i,\mathrm{H}}(\mathbf{f}-\mathbf{f}^{i-1})\right\}  +(\mathbf{f}-\mathbf{f}^{i-1})^{\mathrm{H}}\mathbf{M}_{f}^{i}(\mathbf{f}-\mathbf{f}^{i-1}),\label{eq:f1}
\end{align}
where $\mathbf{m}_{f}^{i}$ and $\mathbf{M}_{f}^{i}$ are to be designed
to satisfy Assumption A.

Assumptions (A1) and (A2) are readily satisfied. Assumption (A4) is
a derivative consistency condition. Denote $\tilde{\mathbf{f}}\in\mathcal{S}_{f}$,
then the directional derivative of $\hat{f}\left(\mathbf{f},\mathbf{f}^{i-1}\right)$
at $\mathbf{f}^{i-1}$ with direction $\tilde{\mathbf{f}}-\mathbf{f}^{i-1}$
is 
\begin{equation}
2\mathrm{Re}\left\{ \mathbf{m}_{f}^{i,\mathrm{H}}(\tilde{\mathbf{f}}-\mathbf{f}^{i-1})\right\} .\label{eq:mq}
\end{equation}
The corresponding directional derivative of $f\left(\mathbf{f}\right)$
is 
\begin{equation}
\frac{-\theta e^{-\theta x^{i}}}{\left(1+e^{-\theta x^{i}}\right)^{2}}2\mathrm{Re}\left\{ \mathbf{f}^{i-1,\mathrm{H}}\mathbf{\mathbf{G}}^{i,\mathrm{H}}\mathbf{e}^{i-1}\mathbf{e}^{i-1,\mathrm{H}}\mathbf{G}\left(\tilde{\mathbf{f}}-\mathbf{f}^{i-1}\right)\right\} ,\label{eq:m2}
\end{equation}
where $x^{i}$ is given in (\ref{eq:Q5}).

Assumption (A4) is satisfied only when (\ref{eq:mq}) and (\ref{eq:m2})
are equal, yielding 
\begin{equation}
\mathbf{m}_{f}^{i}=-\frac{\theta e^{-\theta x^{i}}}{\left(1+e^{-\theta x^{i}}\right)^{2}}\mathbf{\mathbf{G}}^{i,\mathrm{H}}\mathbf{e}^{i-1}\mathbf{e}^{i-1,\mathrm{H}}\mathbf{G}^{i}\mathbf{f}^{i-1}.
\end{equation}

In order for Assumption (A3) to hold, it is sufficient to show that
$\hat{f}(\mathbf{f},\mathbf{f}^{i-1})$ is an upper bound for each
linear cut in any direction. In particular, let $\mathbf{f}=\mathbf{f}^{i-1}+\xi(\tilde{\mathbf{f}}-\mathbf{f}^{i-1}),\forall\xi\in[0,1]$,
we need to show 
\begin{align}
f\left(\mathbf{f}^{i-1}+\xi(\tilde{\mathbf{f}}-\mathbf{f}^{i-1})\right) & \leq f\left(\mathbf{f}^{i-1}\right)+2\xi\mathrm{Re}\left\{ \mathbf{m}_{f}^{i,\mathrm{H}}(\tilde{\mathbf{f}}-\mathbf{f}^{i-1})\right\} 
+  \xi^{2}(\tilde{\mathbf{f}}-\mathbf{f}^{i-1})^{\mathrm{H}}\mathbf{M}_{f}^{i}(\tilde{\mathbf{f}}-\mathbf{f}^{i-1}).\label{line}
\end{align}

Define $L(\xi)=f(\mathbf{f}^{i-1}+\xi(\tilde{\mathbf{f}}-\mathbf{f}^{i-1}))$
and $l(\xi)=\gamma\sigma^{2}-|\mathbf{e}^{i-1,\mathrm{H}}\mathbf{G}^{i}(\mathbf{f}^{i-1}+\xi(\widetilde{\mathbf{f}}-\mathbf{f}^{i-1}))|^{2}$.
(\ref{line}) holds if the second-order derivative of $L(\xi)$ is
no more than that of the right hand side of (\ref{line}). The corresponding
sufficient condition can be formulated as 
\begin{equation}
\frac{\partial^{2}L(\xi)}{\partial\xi^{2}}\leq2\mathrm{\mathrm{Tr}}\left[(\tilde{\mathbf{f}}-\mathbf{f}^{i-1})^{\mathrm{H}}\mathbf{M}_{f}^{i}(\tilde{\mathbf{f}}-\mathbf{f}^{i-1})\right].\label{eq:suffi}
\end{equation}

Before deriving the expression of $\partial^{2}L(\xi)/\partial\xi^{2}$,
we calculate the first-order derivative of $L(\xi)$, as follows 
\begin{align}
\frac{\partial L(\xi)}{\partial\xi} & =g(\xi)\nabla_{\xi}l(\xi),
\end{align}
where $g(\xi)=\frac{\theta e^{-\theta l(\xi)}}{(1+e^{-\theta l(\xi)})^{2}}$,
$\nabla_{\xi}l(\xi)=-2\mathrm{Re}\{\mathbf{q}^{\mathrm{H}}(\widetilde{\mathbf{f}}-\mathbf{f}^{i-1})\}$,
and $\mathbf{q}=\mathbf{\mathbf{G}}^{i,\mathrm{H}}\mathbf{e}^{i-1}\mathbf{e}^{i-1,\mathrm{H}}\mathbf{G}^{i}(\mathbf{f}^{i-1}+\xi(\widetilde{\mathbf{f}}-\mathbf{f}^{i-1}))$.

Then, the second-order derivative is derived as 
\begin{align}
\frac{\partial^{2}L(\xi)}{\partial\xi^{2}} & =g(\xi)\nabla_{\xi}^{2}l(\xi)-\theta g(\xi)\nabla_{\xi}l(\xi)\left(\nabla_{\xi}l(\xi)\right)^{\mathrm{T}} +2\left(1+e^{-\theta l(\xi)}\right)g(\xi)\nabla_{\xi}l(\xi)\left(g(\xi)\nabla_{\xi}l(\xi)\right)^{\mathrm{T}},\label{eq:qua1}
\end{align}
where $\nabla_{\xi}^{2}l(\xi)=-2\mathrm{Re}\{(\widetilde{\mathbf{f}}-\mathbf{f}^{i-1})^{\mathrm{H}}\boldsymbol{\Theta}(\widetilde{\mathbf{f}}-\mathbf{f}^{i-1})\}$
and $\boldsymbol{\Theta}=\xi\mathbf{\mathbf{G}}^{i,\mathrm{H}}\mathbf{e}^{i-1}\mathbf{e}^{i-1,\mathrm{H}}\mathbf{G}^{i}$.

(\ref{eq:qua1}) is rewritten as a quadratic form of $\mathbf{t}=\widetilde{\mathbf{f}}-\mathbf{f}^{i-1}$,
as follows 
\begin{align}
\frac{\partial^{2}L(\xi)}{\partial\xi^{2}} & =\left[\begin{array}{c}
\mathbf{t}\\
\mathbf{t}{}^{*}
\end{array}\right]^{\mathrm{H}}\boldsymbol{\Phi}\left[\begin{array}{c}
\mathbf{t}\\
\mathbf{t}{}^{*}
\end{array}\right],\label{eq:vv}
\end{align}
where 
\begin{align*}
\boldsymbol{\Phi} & =g(\xi)\left(2\left(1+e^{-\theta l(\xi)}\right)g(\xi)-\theta\right)\left[\begin{array}{c}
\mathbf{q}\\
\mathbf{q}^{\mathrm{*}}
\end{array}\right]\left[\begin{array}{c}
\mathbf{q}\\
\mathbf{q}^{\mathrm{*}}
\end{array}\right]^{\mathrm{H}}-g(\xi)\mathbf{I}_{2}\otimes\boldsymbol{\Theta}.
\end{align*}

Furthermore, we also manipulate the right hand side of (\ref{eq:suffi})
into the same form as in (\ref{eq:vv}), as follows 
\begin{align}
\left[\begin{array}{c}
\mathbf{t}\\
\mathbf{t}{}^{*}
\end{array}\right]^{\mathrm{H}}\left[\begin{array}{cc}
\mathbf{I}\otimes\mathbf{M}_{f}^{i} & \mathbf{0}\\
\mathbf{0} & \mathbf{I}\otimes\mathbf{M}_{f}^{i,\mathrm{T}}
\end{array}\right]\left[\begin{array}{c}
\mathbf{t}\\
\mathbf{t}{}^{*}
\end{array}\right].\label{eq:bb}
\end{align}

Combining (\ref{eq:vv}) and (\ref{eq:bb}), the sufficient condition
in (\ref{eq:suffi}) is equivalent to 
\begin{align*}
\left[\begin{array}{c}
\mathbf{t}\\
\mathbf{t}{}^{*}
\end{array}\right]^{\mathrm{H}}\boldsymbol{\Phi}\left[\begin{array}{c}
\mathbf{t}\\
\mathbf{t}{}^{*}
\end{array}\right] & \leq\left[\begin{array}{c}
\mathbf{t}\\
\mathbf{t}{}^{*}
\end{array}\right]^{\mathrm{H}}\left[\begin{array}{cc}
\mathbf{I}\otimes\mathbf{M}_{f}^{i} & \mathbf{0}\\
\mathbf{0} & \mathbf{I}\otimes\mathbf{M}_{f}^{i,\mathrm{T}}
\end{array}\right]\left[\begin{array}{c}
\mathbf{t}\\
\mathbf{t}{}^{*}
\end{array}\right],
\end{align*}
which can be satisfied when $\mathbf{M}_{f}^{i}$ satisfies 
\[
\boldsymbol{\Phi}\preceq\left[\begin{array}{cc}
\mathbf{I}\otimes\mathbf{M}_{f}^{i} & \mathbf{0}\\
\mathbf{0} & \mathbf{I}\otimes\mathbf{M}_{f}^{i,\mathrm{T}}
\end{array}\right].
\]
For convenience, we select $\mathbf{M}_{f}^{i}=\alpha_{f}^{i}\mathbf{I}=\lambda_{\mathrm{\max}}\left(\boldsymbol{\Phi}\right)\mathbf{I}$.
Then, $\hat{f}\left(\mathbf{f},\mathbf{f}^{i-1}\right)$ in (\ref{eq:f1})
is designed to be 
\begin{align*}
 & \hat{f}\left(\mathbf{f},\mathbf{f}^{i-1}\right)\\
 & =f\left(\mathbf{f}^{i-1}\right)+2\mathrm{Re}\left\{ \mathbf{m}_{f}^{i,\mathrm{H}}(\mathbf{f}-\mathbf{f}^{i-1})\right\} +\alpha_{f}^{i}||\mathbf{f}-\mathbf{f}^{i-1}||_{2}^{2}\\
 & =2\textrm{\ensuremath{\mathrm{Re}}}\left\{ \mathbf{d}_{f}^{i,\mathrm{H}}\mathbf{f}\right\} +\alpha_{f}^{i}||\mathbf{f}||_{2}^{2}+\textrm{const}_{f}^{i},
\end{align*}
where $\mathbf{d}_{f}^{i,\mathrm{H}}$, $\alpha_{f}^{i}$ and $\textrm{const}_{f}^{i}$
are defined in Lemma \ref{lemma-f}. The deterministic expression
of $\lambda_{\mathrm{\max}}\left(\boldsymbol{\Phi}\right)$ is difficult
to obtain, therefore we derive the upper bound shown as follows 
\begin{align*}
 & \lambda_{\mathrm{\max}}\left(\boldsymbol{\Phi}\right)\\
 & \overset{\mathrm{\textrm{(p1)}}}{\leq}2\left(1+e^{-\theta l(\xi)}\right)g^{2}(\xi)\lambda_{\mathrm{max}}\left(\left[\begin{array}{c}
\mathbf{q}\\
\mathbf{q}^{\mathrm{*}}
\end{array}\right]\left[\begin{array}{c}
\mathbf{q}\\
\mathbf{q}^{\mathrm{*}}
\end{array}\right]^{\mathrm{H}}\right)\\
 & ~~~~~-g(\xi)\lambda_{\mathrm{\min}}\left(\mathbf{I}_{2}\otimes\boldsymbol{\Theta}\right)-\theta g(\xi)\lambda_{\mathrm{min}}\left(\left[\begin{array}{c}
\mathbf{q}\\
\mathbf{q}^{\mathrm{*}}
\end{array}\right]\left[\begin{array}{c}
\mathbf{q}\\
\mathbf{q}^{\mathrm{*}}
\end{array}\right]^{\mathrm{H}}\right)\\
 & \overset{\mathrm{\textrm{(p2)}}}{=}4\left(1+e^{-\theta l(\xi)}\right)g^{2}(\xi)||\mathbf{q}||_{2}^{2}\\
 & \overset{\mathrm{\textrm{(p3)}}}{<}\frac{\theta^{2}}{2}||\mathbf{q}||_{2}^{2}\\
 & \overset{\mathrm{\textrm{(p4)}}}{\leq}\frac{\theta^{2}}{2}\lambda_{\mathrm{max}}(\mathbf{\mathbf{G}}^{i,\mathrm{H}}\mathbf{e}^{i-1}\mathbf{e}^{i-1,\mathrm{H}}\mathbf{G}^{i}\mathbf{\mathbf{G}}^{i,\mathrm{H}}\mathbf{e}^{i-1}\mathbf{e}^{i-1,\mathrm{H}}\mathbf{G}^{i})\cdot||\mathbf{f}^{i-1}+\gamma(\widetilde{\mathbf{f}}-\mathbf{f}^{i-1})||_{2}^{2}\\
 & \overset{\textrm{\textrm{\ensuremath{\textrm{(p5)}}}}}{\leq}\frac{\theta^{2}}{2}P_{max}|\mathbf{e}^{i-1,\mathrm{H}}\mathbf{G}^{i}\mathbf{\mathbf{G}}^{i,\mathrm{H}}\mathbf{e}^{i-1}|^{2}.
\end{align*}
The above inequalities are due to the following mathematical properties:

(p1): $\mathbf{A}$ and $\mathbf{B}$ are Hermitian matrices: $\lambda_{\mathrm{\max}}(\mathbf{A})+\lambda_{\mathrm{\max}}(\mathbf{B})\geq\lambda_{\mathrm{\max}}(\mathbf{A}+\mathbf{B})$
\cite{maher1999handbook}.

(p2): $\mathbf{A}$ is rank one: $\lambda_{\mathrm{max}}(\mathbf{A})=\mathrm{\mathrm{Tr}}\left[\mathbf{A}\right],\lambda_{\mathrm{min}}(\mathbf{A})=0$
\cite{maher1999handbook}.

(p3): $\left(1+e^{-\theta l(\xi)}\right)g^{2}(\xi)\leq\theta^{2}/8$,
where the equality holds when $l(\xi)=0$.

(p4): $\mathbf{A}$ is positive semidefinite with maximum eigenvalue
$\lambda_{\mathrm{max}}(\mathbf{A})$ and $\mathbf{B}$ is positive
semidefinite: $\mathrm{\mathrm{Tr}}\left[\mathbf{A}\mathbf{B}\right]\leq\lambda_{\mathrm{max}}(\mathbf{A})\mathrm{\mathrm{Tr}}\left[\mathbf{B}\right]$
\cite{maher1999handbook}.

(p5): Power constraint: $||\mathbf{f}^{i-1}+\gamma(\widetilde{\mathbf{f}}-\mathbf{f}^{i-1})||_{2}^{2}\leq P_{max}$.

Hence, the proof is completed.

\section{The proof of Theorem \ref{theorem-1}\label{subsec:The-proof-2}}

Define the random functions 
\begin{align}
g^{n}(\mathbf{x}) & =\frac{1}{n}\sum_{i=1}^{n}f\left(\mathbf{x}|\mathbf{G}^{i}\right),\label{eq:d3}\\
\hat{g}^{n}(\mathbf{x}) & =\frac{1}{n}\sum_{i=1}^{n}\hat{f}\left(\mathbf{x},\mathbf{x}^{i-1}|\mathbf{G}^{i}\right).\label{eq:d4}
\end{align}

To state the convergence result, we need the following lemmas.

\begin{lemma}\label{Lemma3} Suppose Assumptions A and B are satisfied
and define a limit point $\bar{\mathbf{x}}$ of a subsequence $\{\mathbf{x}^{n_{j}}\}_{j=1}^{\infty}$,
then there exists uniformly continuous functions $g(\mathbf{x})$
and $\hat{g}(\mathbf{x})$ such that 
\begin{align}
g(\mathbf{x}) & =\lim_{n\rightarrow\infty}g^{n}(\mathbf{x})=\mathbb{E}\left[f\left(\mathbf{x}|\mathbf{G}\right)\right],\forall\mathbf{x}\in\mathcal{S}_{x},\label{eq:d5}\\
g(\bar{\mathbf{x}}) & =\lim_{j\rightarrow\infty}g^{n_{j}}(\mathbf{x}^{n_{j}}),\label{eq:d6}\\
\hat{g}(\mathbf{x}) & =\lim_{j\rightarrow\infty}\hat{g}^{n_{j}}(\mathbf{x}),\forall\mathbf{x}\in\mathcal{S}_{x},\label{eq:d1}\\
\hat{g}(\bar{\mathbf{x}}) & =\lim_{j\rightarrow\infty}\hat{g}^{n_{j}}(\mathbf{x}^{n_{j}}).\label{eq:d2}
\end{align}
\end{lemma}

\textbf{\textit{Proof:}} First, it follows that $f\left(\mathbf{x},\mathbf{G}\right)$
is bounded for $\forall\mathbf{x}\in\mathcal{S}_{x}$ and all channel
realizations due to the assumption (B2), and therefore, (\ref{eq:d5})
holds by using the strong law of large numbers \cite{large-number-law}.
Also, the families of functions $\{g^{n_{j}}(\mathbf{x})\}$ are equicontinuous
and bounded over a compact set $\mathcal{S}_{x}$ due to the assumption
(B2) and the use of mean value theorem. Thus, by restricting to a
subsequence, we have (\ref{eq:d6}). Furthermore, the families of
functions $\{\hat{g}^{n}(\mathbf{x})\}$ are also equicontinuous and
bounded over a compact set $\mathcal{S}_{x}$ due to the assumption
(B2) that $||\nabla_{\mathbf{x}}\hat{f}(\mathbf{x},\mathbf{x}^{i-1},\mathbf{G})||$
is bounded. Hence the Arzelà-Ascoli theorem \cite{A-Atheorem} implies
that, by restricting to a subsequence, there exists a uniformly continuous
function $\hat{g}(\mathbf{x})$ such that (\ref{eq:d1}) and (\ref{eq:d2})
hold.\hspace{15.2cm}$\blacksquare$

On the other hand, the update rule of Algorithm 1 leads the following
lemma.

\begin{lemma}\label{Lemma4}$\lim_{n\rightarrow\infty}|\hat{g}^{n}(\mathbf{x}^{n})-g^{n}(\mathbf{x}^{n})|=0$,
almost surely. \end{lemma}

\textbf{\textit{Proof:}} The proof of Lemma \ref{Lemma4} is the same
with that of (\cite{SSM}, Lemma 1) and is omitted for conciseness.\hspace{13.9cm}$\blacksquare$

Assumption (A3) implies that $\hat{g}^{n_{j}}(\mathbf{x})\geq g^{n_{j}}(\mathbf{x}),\forall\mathbf{x}\in\mathcal{S}_{x}$.
combining it with (\ref{eq:d5}) and (\ref{eq:d1}), we obtain 
\begin{equation}
\hat{g}(\mathbf{x})\geq g(\mathbf{x}),\forall\mathbf{x}\in\mathcal{S}_{x}.\label{eq:d7}
\end{equation}

Moreover, combining Lemma \ref{Lemma4} with (\ref{eq:d6}) and (\ref{eq:d2}),
it yields 
\begin{equation}
\hat{g}(\bar{\mathbf{x}})=g(\bar{\mathbf{x}}).\label{eq:d8}
\end{equation}

Then, (\ref{eq:d7}) and (\ref{eq:d8}) imply that $\bar{\mathbf{x}}$
is a minimizer of function $\hat{g}(\mathbf{x})-g(\mathbf{x})$, hence
the first-order optimality condition satisfies 
\begin{equation}
\nabla\hat{g}(\bar{\mathbf{x}})-\nabla g(\bar{\mathbf{x}})=0.\label{eq:d9}
\end{equation}

Due to the fact that $\bar{\mathbf{x}}$ is the limit point of Problem
(\ref{Pro:singlef}) or Problem (\ref{Pro:singlee}), we have $\hat{g}(\bar{\mathbf{x}})\leq\hat{g}(\mathbf{x}),\forall\mathbf{x}\in\mathcal{S}_{x}$,
which implies that 
\begin{equation}
\left\langle \nabla\hat{g}(\bar{\mathbf{x}}),\mathbf{x}-\bar{\mathbf{x}}\right\rangle \geq0,,\forall\mathbf{x}\in\mathcal{S}_{x}.
\end{equation}

Combining this with (\ref{eq:d9}), we obtain 
\begin{equation}
\left\langle \nabla g(\bar{\mathbf{x}}),\mathbf{x}-\bar{\mathbf{x}}\right\rangle \geq0,,\forall\mathbf{x}\in\mathcal{S}_{x},\label{eq:vvv}
\end{equation}
which means that the directional derivative of the objective function
$g(\mathbf{x})$ is non-negative for every feasible direction at $\bar{\mathbf{x}}$.
Recalling that $\mathbf{x}\in\{\mathbf{f},\mathbf{e}\}$ and defining
the limit points $\{\bar{\mathbf{f}},\bar{\mathbf{e}}\}$, (\ref{eq:vvv})
is equivalent to 
\begin{align*}
\begin{cases}
\left\langle \nabla g(\bar{\mathbf{f}}),\mathbf{f}-\bar{\mathbf{f}}\right\rangle \geq0,\forall\mathbf{f}\in\mathcal{S}_{f},\\
\left\langle \nabla g(\bar{\mathbf{e}}),\mathbf{e}-\bar{\mathbf{e}}\right\rangle \geq0,\forall\mathbf{e}\in\mathcal{S}_{e}.
\end{cases}
\end{align*}

Therefore, according to \cite{thesis-Razaviyayn}, $\{\bar{\mathbf{f}},\bar{\mathbf{e}}\}$
is a stationary point of Problem (\ref{Pro:single2}) due to the regularity
of $g(\cdot)$.

\section{The proof of Theorem \ref{theorem-1}\label{subsec:The-proof-4}}

Define the random functions 
\begin{align}
G^{n}(\mathbf{x}) & =\frac{1}{n}\sum_{i=1}^{n}F\left(\mathbf{x}|\mathbf{G}^{i}\right),\label{eq:d3-1}\\
\hat{G}^{n}(\mathbf{x}) & =\frac{1}{n}\sum_{i=1}^{n}\hat{F}\left(\mathbf{x},\mathbf{x}^{i-1}|\mathbf{G}^{i}\right).\label{eq:d4-1}
\end{align}

To state the convergence result, we need the following lemmas.

\begin{lemma}\label{Lemma5} Suppose Assumptions B and C are satisfied
and define a limit point $\bar{\mathbf{x}}$ of a subsequence $\{\mathbf{x}^{n_{j}}\}_{j=1}^{\infty}$,
then there exists uniformly continuous functions $G(\mathbf{x})$
and $\hat{G}(\mathbf{x})$ such that 
\begin{align}
G(\mathbf{x}) & =\lim_{n\rightarrow\infty}G^{n}(\mathbf{x})=\mathbb{E}\left[F\left(\mathbf{x}|\mathbf{G}\right)\right],\forall\mathbf{x}\in\mathcal{S}_{x},\label{eq:d5-1}\\
G(\bar{\mathbf{x}}) & =\lim_{j\rightarrow\infty}G^{n_{j}}(\mathbf{x}^{n_{j}}),\label{eq:d6-1}\\
\hat{G}(\mathbf{x}) & =\lim_{j\rightarrow\infty}\hat{G}^{n_{j}}(\mathbf{x}),\forall\mathbf{x}\in\mathcal{S}_{x},\label{eq:d1-1}\\
\hat{G}(\bar{\mathbf{x}}) & =\lim_{j\rightarrow\infty}\hat{G}^{n_{j}}(\mathbf{x}^{n_{j}}).\label{eq:d2-1}
\end{align}
\end{lemma}

\textbf{\textit{Proof:}} The proof of Lemma \ref{Lemma5} is the same
with that of Lemma \ref{Lemma3} and is omitted for brevity.\hspace{0.1cm}$\blacksquare$

On the other hand, $\mathbf{x}^{n_{j}}$ is the minimizer of $\hat{G}^{n_{j}}(\mathbf{x})$,
thus 
\begin{equation}
\hat{G}^{n_{j}}(\mathbf{x}^{n_{j}})\leq\hat{G}^{n_{j}}(\mathbf{x}),\forall\mathbf{x}\in\mathcal{S}_{x}.
\end{equation}

Assuming $j\rightarrow\infty$, and combining (\ref{eq:d1-1}) and
(\ref{eq:d2-1}), we obtain $\hat{G}(\bar{\mathbf{x}})\leq\hat{G}(\mathbf{x}),\forall\mathbf{x}\in\mathcal{S}_{x}$,
which implies that its first-order optimality condition satisfies
\begin{equation}
\left\langle \nabla\hat{G}(\bar{\mathbf{x}}),\mathbf{x}-\bar{\mathbf{x}}\right\rangle \geq0,,\forall\mathbf{x}\in\mathcal{S}_{x}.\label{cx}
\end{equation}

By combining (\ref{cx}) and Assumption (C3), we finally obtain 
\begin{equation}
\left\langle \nabla G(\bar{\mathbf{x}}),\mathbf{x}-\bar{\mathbf{x}}\right\rangle \geq0,,\forall\mathbf{x}\in\mathcal{S}_{x}.\label{cx-1}
\end{equation}

Since that $\mathbf{x}\in\{\mathbf{F},\mathbf{e}\}$, we define the
limit points $\{\bar{\mathbf{F}},\bar{\mathbf{e}}\}$, (\ref{cx-1})
is then equivalent to 
\begin{align*}
\begin{cases}
\left\langle \nabla G(\bar{\mathbf{F}}),\mathbf{F}-\bar{\mathbf{F}}\right\rangle \geq0,\forall\mathbf{F}\in\mathcal{S}_{f},\\
\left\langle \nabla G(\bar{\mathbf{e}}),\mathbf{e}-\bar{\mathbf{e}}\right\rangle \geq0,\forall\mathbf{e}\in\mathcal{S}_{e}.
\end{cases}
\end{align*}

Therefore, according to \cite{thesis-Razaviyayn}, $\{\bar{\mathbf{F}},\bar{\mathbf{e}}\}$
is a stationary point of Problem (\ref{Pro:multi2}) due to the regularity
of $G(\cdot)$.

 \bibliographystyle{IEEEtran}
\bibliography{bibfile}

\end{document}